\documentstyle[12pt,epsfig]{article}

\newcommand{\PT}{p_{{T}}}
\newcommand{\AN}{A_{{N}}}

\newcommand{\FPT}{{F}}
\newcommand{\DRF}{{D}}
\newcommand{\GXA}{{G}}
\newcommand{\IA}{{{a}}}
\newcommand{\IB}{{{b}}}
\newcommand{\IC}{{{c}}}

\newcommand{\Pol}{\bf {P} }
\begin{document}
\begin{flushright}
 Preprint IHEP 2001-13\\
 hep-ph/0111128
\end{flushright}
\vskip 1cm
\centerline{ \bf\Large {Universal scaling behaviour of the transverse}}
\centerline{ \bf\Large {polarization for inclusively produced hyperons}}
\centerline{ \bf\Large {in hadron-hadron collisions}}
\begin{center} V.V.Abramov$^{1}$
\end{center}
\vskip 0.5cm
\centerline { Experimental Physics Department,}
\centerline { Institute for High Energy Physics, P.O. Box 35, }
\centerline{ Protvino, 142281 Moscow region, Russia}
\centerline {$^{1}$E-mail: abramov\_v@mx.ihep.su}
\begin{abstract}
    Experimental data on the  polarization of hyperons, inclusively  produced
in hadron-hadron collisions,   have been analyzed.
 It is shown that the existing  data  can be described  by a function of 
  transverse momentum  ($p_{T}$)  and  
two scaling  variables  $x_{A\pm}$ = $(x_{R} \pm x_{F})/2$: 
$ {\Pol_{H}} = A^{\alpha} {\FPT}(p_{T})[{\GXA}(x_{A+} -x_{2}) - 
\sigma {\GXA}(x_{A-} +x_{2}) ]$.
The  function ${\GXA}(x_{A+})$ is proportional to  
$ \sin[\omega(x_{A+} -x_{1})]/\omega$, which results for some reactions in an
oscillation of ${\Pol_{H}}$ as a function of $x_{F}$.
 The $\omega$, as well as the magnitude and the sign of the hyperon 
polarization depend on quark composition of hadrons participating in 
a reaction.  The atomic weight dependence of 
the $\Lambda$ hyperon polarization is characterized by
 the parameter  $\alpha \approx -0.16|x_{F}|$. There is an analogy between the
 scaling properties of  the hyperon polarization and the analyzing power
 ($\AN$) in hadron production reactions.  This new scaling law  allows one 
to predict hyperon polarization  for reactions and kinematic regions, yet
 unexplored  in experiments and to confront these predictions with future
 experiments and  various models.
\end{abstract}


%
\newpage
\section{Introduction}
 The understanding of spin-dependent effects in inclusive hadron
production processes in the framework of QCD is still far from being
satisfactory, despite significant experimental and theoretical progress 
over the past few years. In particular, the study of hyperon polarization
($\Pol_{H}$) and the analyzing power ($\AN$) could provide invaluable and
completely new insight into the field of ``spin physics'' and, in addition, 
might also yield a better understanding of the hadronization process.

  In this paper we will study  the existing data for one measured 
spin-dependent quantity (transverse hyperon polarization $\Pol_{H}$ in 
inclusive reactions $ a + b\rightarrow c^{\uparrow} + X$ ) from an empirical
 point of view in collisions of unpolarized protons, antiprotons, 
$K^{\pm}$, $\pi^{\pm}$  or hyperons  with protons or nuclei.

    Experiments on hyperon production performed during more than two  decades
 since the first  polarization observation \cite{Bunce}
 have shown that the hyperon polarization is significant in a wide range
 of beam energies. 
  Almost all the existing data (with an equivalent proton momentum 
on a fixed target from 4 up to 2049 GeV/c) at
medium and high energies are used for the analysis. 

 The study of $\Lambda$ hyperon polarization in $pp$ and $pA$ collisions
 has been carried out in many experiments and  revealed an approximate
scaling behaviour as a function of $x_{F}$ variable at fixed $p_{T}$, as well
 as a function of $p_{T}$ at fixed $x_{F}$ \cite{Abe_D34}-\cite{Smith}. 
Further investigations of the hyperon and proton polarization using different
 beams and targets have shown its dependence on the hyperon and the beam 
flavors, as well
 as on the target atomic weight  \cite{Morelos}-\cite{Polvado}. It has 
to be mentioned that a large value of the hyperon polarization represents
a significant problem for the existing strong interaction theory 
which predicts, in the framework of  pertubative Quantum Chromodynamics (pQCD),
 the vanishing of the polarization at high $p_{T}$ and energy \cite{Kane}.
For the review of the hyperon polarization data and the existing models
 see \cite{Pondrom,Lach,Soffer}.

Recently a new scaling law has been proposed for a hadron production  
analyzing power ($\AN$) in reactions
$$   a^{\uparrow} +  b\rightarrow c + X $$
where $a$, $b$ and $c$ are hadrons, and the hadron $a$ is transversely 
polarized \cite{Prep98,EPJC}.
Since the hyperon polarization and the hadron analyzing power could be closely
 related \cite{Liang1}, it is reasonable to expect that similar  scaling 
properties  take place for the inclusive hyperon production in collisions of 
 unpolarized  hadron beams with protons or nuclei. According to  Ref. 
\cite{EPJC} $\AN$ at high energies ($\ge 40$ GeV) and high transverse momentum
 $p_{T}$ ($\ge 1$ GeV/c) can be described  by a simple function of two 
variables, $p_{T}$ and $x_{A}$:
\begin{equation}
  \AN = {\FPT}(p_{T}){\GXA}(x_{A}),    
\label{eq:FGX}
\end{equation}
where 
\begin{equation}
     x_{A} = p_{\IC{}}\cdot p_{\IB}/p_{\IA}\cdot p_{\IB}= E_{c}/E_{a}.
\label{eq:XA}
\end{equation}
Here, $p_{\IC}$, $p_{\IB}$ and $p_{\IA}$ are four-momenta of the produced
hadron, the target hadron and the beam hadron, respectively. The energies
$E_{c}$ (produced hadron) and $E_{a}$ (beam hadron) are measured in 
a reference frame, where a polarized ($p$ or $\bar{p}\!$) particle strikes 
an unpolarized target ($p$ or $A$) which is at rest. There are also
alternative expressions for the scaling variable $x_{A}$, which are close to
(\ref{eq:XA}) numerically at the high beam and secondary hadron 
energies \cite{EPJC}.  In particular, neglecting masses in (\ref{eq:XA}), we
have in the c.m. frame
\begin{equation}
  x_{A}     = (x_{F}+x_{R})/2, \quad
\label{eq:XFR}
\end{equation}
where $x_{F}=p^{*}_{z}/p^{*}_{max}$ and $x_{R}=p^{*}/p^{*}_{max}$.
Here $p^{*}$, $p^{*}_{z}$ and $p^{*}_{max}$ is momentum of the
produced hadron, its longitudinal component and the maximum possible value
of it, respectively, all in the c.m. reference frame.
Eqs. (\ref{eq:FGX}) and (\ref{eq:XFR})  describe  $\AN$ for most 
of the hadron production reactions almost
 as well as  eqs. (\ref{eq:FGX}) and (\ref{eq:XA}) \cite{EPJC}.
 The only exception is the $\AN$ for $\Lambda$ hyperon production in 
$p^{\uparrow} \!p(A)$ collisions, for which the
variable (\ref{eq:XFR}) is preferable since it gives a smaller $\chi^{2}$
($\chi^{2}/N_{DOF}= 24.3/44$ for (\ref{eq:XFR}) vs 39.4/44 for (\ref{eq:XA})),
where $N_{DOF}$ is the number of degrees of freedom in a fit. It is interesting
 that $\Lambda$ hyperon polarization is also better described by 
eq. (\ref{eq:FGX}) if $x_{A}$ is given by (\ref{eq:XFR}). 

It has to be mentioned that the variable $x=E_{c}/E_{a}$ was used to describe
the scaling properties of the secondary hadron ($\pi, K, \bar{p}, \bar{n}$)
spectra in the region $E_{c}/E_{a} \ge 0.6 \div 0.7$ \cite{Bushnin}.

 A natural generalization of (\ref{eq:XFR}) is given by a linear function
of $x_{R}$ and $x_{F}$ variables with relative weights which could 
be determined from the data fit:
\begin{equation}
x_{A\pm}=x_{R}\cdot w_{1} \pm x_{F} \cdot w_{2},
\label{eq:XFRA}
\end{equation}
where $w_{1} + w_{2} = 1$. For $w_{1} = 0.5$
the variable $x_{A+}$ in (\ref{eq:XFRA}) is close to $|x_{F}|$  in the beam 
fragmentation region  ($x_{F} \approx 1$)   and 
$x_{A+} \approx x_{T}/2$ in the central region ($x_{F} \approx 0$). 
 The variable  $x_{A-}$ have  similar properties in the backward hemisphere
 ($x_{F} \le 0$). For $w_{1} = 0.5$  variable $x_{A+}$ is identical to  
$x_{A}$ (eq.(\ref{eq:XFR})).

The hyperon polarization features are different in some respects from 
the analyzing power features listed in \cite{EPJC}. In particular, we have
 to take into account that the polarization of hyperons produced in $pp$ 
collisions is antisymmetric in $x_{F}$
by virtue of rotational invariance \cite{Lundberg,Felix}:
\begin{equation}
 {\Pol_{H}} (x_{F},p_{T}) = -{\Pol_{H}} (-x_{F},p_{T})
\label{eq:Pasym}
\end{equation}
From the  relation (\ref{eq:Pasym}) we have also for the hyperon 
polarization ${\Pol_{H}} (0,p_{T}) = 0$ for $pp$ or other symmetric
initial states in contrast to the analyzing power, 
which could be different from zero even at $x_{F}=0$, see \cite{EPJC}.
The initial state $p^{\uparrow} \!p$ is not symmetric with respect to
the $180^{o}$ rotation since one proton is polarized.

From a dimensional analysis alone, the polarization ${\Pol_{H}}$ admits two 
types of contributions, which are functions of $u/s$ and $t/s$:
\begin{equation}
   {\Pol_{H}} = 
{\FPT}(p_{T})\Bigl[ {\GXA}_{+}(-u/s) - {\GXA}_{-}(-t/s) \Bigr],
\label{eq:GUT}
\end{equation}
where $u$, $t$ and $s$ are Mandelstam variables:
\begin{eqnarray}
s &=& (p_{a} + p_{b})^{2} \approx +2p_{a}\cdot p_{b}, \nonumber \\
t &=& (p_{a} - p_{c})^{2} \approx -2p_{a}\cdot p_{c}, \\ \label{eq:STU}
u &=& (p_{b} - p_{c})^{2} \approx -2p_{b}\cdot p_{c}, \nonumber
\end{eqnarray}
where in the last column the hadron masses are neglected. It is easy to 
show that at high energies the following approximations are valid:
\begin{equation}
 -u/s  \approx  { (x_{R} + x_{F})/{2} } =  x_{A+},  \\
\label{eq:XAD1EF}
\end{equation}
\begin{equation}
 -t/s  \approx  { (x_{R} - x_{F})/{2} } =  x_{A-}, \\
\label{eq:XAD2EF}
\end{equation}
where $x_{A\pm}$ are given by eq. (\ref{eq:XFRA}) with $w_{1} = 0.5$.
  Relations, similar to (\ref{eq:XAD1EF})-(\ref{eq:XAD2EF}),  
have been used in  Refs. \cite{CIM1,Qui}.
Using eqs.   (7) - (\ref{eq:XAD2EF}) it is easy to show that
$x_{A+} \approx  p_{\IC{}}\cdot p_{\IB}/p_{\IA}\cdot p_{\IB} =x_{A}$,
where $x_{A}$ is given by eq. (\ref{eq:XA}).

The expression, which takes into account (\ref{eq:Pasym})-(\ref{eq:XAD2EF}) 
and other features of $\Pol_{H}$ is given below:
\begin{equation}
  {\Pol_{H}} = A^{\alpha} {\FPT}(p_{T})[ {\GXA}(x_{A+} -x_{2}) 
                             - \sigma  {\GXA}(x_{A-} +x_{2}) ],
\label{eq:Pol}
\end{equation}
where an additional phase $x_{2}$  and a signature parameter $\sigma$
are introduced to take into account a possible
 violation of (\ref{eq:Pasym}) for collisions, different from $pp$ one
(for $pp$ collisions $x_{2} \equiv 0$ and $\sigma \equiv 1$).
It is assumed below that $\sigma=1$ for  $pA$ collisions, while
$x_{2}$ could be different from zero.
The functions ${\FPT}(p_{T})$ and ${\GXA}(x_{A\pm})$ are determined below 
from the experimental data.

Exactly the same approach can be used to derive an expression for 
the ${\AN}$. The difference is that in the last case 
${\GXA}_{+}(x) \neq {\GXA}_{-}(x)$ in (\ref{eq:GUT}), and ${\GXA}_{+}(x_{A+})$
 dominates, at least for the $p^{\uparrow} \!p \rightarrow \pi^{+} + X$ 
reaction, as the data fit show, over ${\GXA}_{-}(x_{A-})$.

 For the process $pp(A) \rightarrow \Lambda + X$ we use an expression
\begin{equation}
    {\FPT}(\PT{})  = 
  1 -  e^{-\kappa p_{T}^{3}} 
\label{eq:F1}
\end{equation}
for the ${\FPT}(\PT{})$, where
 $\kappa $ is a   fit parameter and $\PT$
is measured in GeV/c. The other processes may require a different expression.
 The exact shape of the ${\FPT}(\PT{})$ should be measured in future
 experiments. For the ${\GXA}(x_{A\pm})$ we use an expression
\begin{equation}
  {\GXA}(x_{A\pm}) = {c\over{2\omega}} \cdot \sin [\omega(x_{A\pm} - x_{1}) ],
\label{eq:GX}
\end{equation}
similar to the one, used in \cite{EPJC},
which allows one to reproduce such features, as an approximate linear 
dependence  of $\Pol_{\Lambda}$ on the $x_{F}$, and non-linear, sometimes 
oscillating behaviour of the hyperon polarization for other reactions.
The difference of eq. (\ref{eq:GX}) from the corresponding expression
in \cite{EPJC} is that the former has an additional factor $1/(2\omega)$,
that makes parameters $c$ and $\omega$ less correlated and reflects 
the tendency of hyperon polarization magnitude decrease with $\omega$ rise.
The parameters  $w_{1}$, $c$, $x_{1}$, $x_{2}$, $\sigma$ and $\omega$  are 
determined  from the data.

The $\Lambda$ polarization data fits indicate that both parameters $x_{1}$ 
and $x_{2}$ can depend on $p_{T}$ and only at high enough 
$p_{T} \ge 1.5 $ GeV/c are compatible with $p_{T}$-independent constant 
values:
\begin{equation}
  x_{1} = \eta_{0} - \eta_{1} e^{ -\delta p_{T}^{3}},
\label{eq:x1}
\end{equation}
\begin{equation}
  x_{2} = (1-Z/A)(\xi_{0} - \xi_{1} e^{ -\nu p_{T}^{3} }),
\label{eq:x2}
\end{equation}
where the target dependent factor $(1-Z/A)$ is introduced to insure $x_{2}=0$
  for $pp$ collisions, as required by relation (\ref{eq:Pasym}). For
a neutron target $Z/A=0$, so the phase $x_{2}$ is the only parameter
in the above equations which makes  the hyperon polarization for the proton 
target different from that of  the neutron one. This feature of 
eq. (\ref{eq:x2}) can be used to estimate the difference in the polarization 
of hyperons produced on the proton and the neutron targets.

 The factor $A^{\alpha}$ takes into account
a possible atomic weight dependence of the hyperon polarization. 
The $\alpha$ could be a constant or a function of kinematic variables:
\begin{equation}
 \alpha = \alpha_{1}|x_{F}| + \alpha_{0},
\label{eq:alpha}
\end{equation}
where $\alpha_{0}$ and $\alpha_{1}$ are  fit parameters. 

In the beam fragmentation region $x_{R} \approx x_{F}$, and 
$x_{A+} \approx x_{F}$, while $x_{A-} \approx 0$. So, in this region
 the dependence of eq. (\ref{eq:Pol}) on $x_{F}$ is determined, mainly, by 
the first term ${\GXA}(x_{A+} -x_{2}) \approx  {\GXA}(x_{F}-x_{2})$. 
The second term  in (\ref{eq:Pol}), ${\GXA}(x_{A-} +x_{2}) \approx 
{\GXA}(x_{2})$ has a weak $x_{F}$ dependence.
 In the target fragmentation region $x_{F}$ dependence is determined, mainly,
 by the $\sigma {\GXA}(x_{A-} +x_{2}) \approx \sigma {\GXA}(-x_{F} +x_{2})$.

Eq. (\ref{eq:Pol}) for $\sigma=\pm 1$ can be expressed using (\ref{eq:GX})
in a different way, with the explicit $x_{R}$ and $x_{F}$ dependences:
\begin{equation}
  {\Pol_{H}} = {c\over{\omega}} A^{\alpha}\cdot {\FPT}(p_{T}) \cases 
{ \cos [\omega(x_{R}w_{1} - x_{1}) ] \sin [\omega(x_{F}w_{2} - x_{2}) ],
 & if $\sigma = +1$; \cr
  \sin [\omega(x_{R}w_{1} - x_{1}) ] \cos [\omega(x_{F}w_{2} - x_{2}) ],
 & if $\sigma = -1$. \cr}
\label{eq:Pol1}
\end{equation}
 Eq. (\ref{eq:Pol}) has a more general form, but it
coincides with (\ref{eq:Pol1}) if we choose eq. (\ref{eq:GX}) for
${\GXA}(x_{A\pm})$ and set $\sigma=\pm 1$. The magnitude of the hyperon
polarization is about $cA^{\alpha}/\omega$. It is assumed here that 
${\FPT}(p_{T})$ is equal unity at its maximum.

In case of the analyzing power measurements we have a non-zero contribution to
the left-right asymmetry from the polarized hadron (beam or target) only and 
the feature (\ref{eq:Pasym}) is not valid. Naively, we may say that  the 
first term, ${\GXA}(x_{A+} -x_{2})$, in (\ref{eq:Pol}), corresponding to 
a polarized beam, gives the main contribution to the analyzing power
 and, as a result, we have eq. (\ref{eq:FGX}) for it. For the case of $\Lambda$
hyperon polarization in the $pp$ collisions both terms in (\ref{eq:Pol}), 
corresponding to the beam fragmentation and the target fragmentation,
 have non-zero contributions and cancel each  other at $x_{F}=0$, in 
accordance to (\ref{eq:Pasym}). 

Relation (\ref{eq:Pasym}) is not valid in general for collisions of 
different hadrons, like $K^{-}p(A)$ or $\bar{p}p$ collisions. In these cases 
we have to use different
functions ${\GXA}_{+}(x_{A+})$ and  ${\GXA}_{-}(x_{A-})$ of $x_{A+}$ 
and $x_{A-}$, respectively. In particular,  parameter $x_{2}$ can be different
from zero and $\sigma$ can be different from unity.

It should be also mentioned, that in the case of  a linear function 
${\GXA}(x_{A\pm})$  and $\sigma=1$ the $x_{R}$ terms are cancelled in 
(\ref{eq:Pol}). The use of $\sin(x)$ in (\ref{eq:GX}) makes
  ${\GXA}(x_{A\pm})$ non-linear,  that prevents complete cancellation 
of $x_{R}$ terms in (\ref{eq:Pol}). A linear case corresponds to the limit
$\omega \rightarrow 0$ in the above equations, while the data fits give
for $pp(A) \rightarrow \Lambda + X$ process  $\omega \approx 3$,
indicating a rather non-linear behaviour of ${\GXA}(x_{A\pm})$ at large
values of arguments. As we will see in the following sections the $\omega$
can be expressed for different  reactions via a linear combination of
terms, which are functions of quantum numbers characterizing the particular 
reaction (see eq. (\ref{eq:omega})). Eq. (\ref{eq:omega})
is used below to fix the $\omega$ in eq. (\ref{eq:Pol}) in data fits that 
allows us to estimate the other parameters with a better accuracy.

  It follows from  eqs. (\ref{eq:Pol}) - (\ref{eq:alpha}) that for the
hadron-proton collisions the hyperon polarization $\Pol_{H}$ and the analyzing
power $\AN$ obey the Helmholtz equation:
\begin{equation}
 {\partial ^{2}{\Pol_{H}} \over \partial x_{A+}^{2}} + 
 {\partial ^{2}{\Pol_{H}} \over \partial x_{A-}^{2}} + 
\omega^{2} {\Pol_{H}} =0.
\label{eq:Helm}
\end{equation}
Eq. (\ref{eq:Helm}) reflects the non-linearity of the $\Pol_{H}$ dependence
and its oscillation as a function of  scaling variables $x_{A\pm}$, which is 
characterized by the $\omega$ parameter.

 Most of the hyperon production experiments have been performed on nuclear
targets, where  relation (\ref{eq:Pasym}) may not be exactly valid.
 Also, practically all the available data are concentrated in the forward 
hemisphere. We cannot exclude that future measurements in the backward 
hemisphere (nuclear target fragmentation region) will show some deviation 
from (\ref{eq:alpha}), which will require a correction of the polarization
$A$-dependence. An estimate of nuclear effects in the forward hemisphere will 
be given below on the base of existing data. We assume also for simplicity that
the $\omega$ parameter is the same for the forward and the backward hemispheres
and reflects the dynamics of strong interaction.

\section{ Lambda hyperon polarization in $pp$ and $pA$ collisions}
The  polarization of ${\Lambda}$ hyperons ($\Pol_{\Lambda}$)
in $pp(A)$ collisions  has been measured in many experiments 
 \cite{Abe_D34}-\cite{Smith}. The data  include proton collisions with
 protons and different nuclei ($d, Be, Cu, W$ and $Pb$).

  We assume here that for the positive polarization and scattering to the left
the hyperon spin is directed along the unit vector 
$\overrightarrow{n} \equiv \overrightarrow{p_{a}} \times \overrightarrow{p_{c}}
/|\overrightarrow{p_{a}} \times \overrightarrow{p_{c}}|$, which is the normal 
to the production plane. Here, $\overrightarrow{p_{a}}$ and
 $\overrightarrow{p_{c}}$ is
a momentum of the beam hadron and that of the produced hyperon, respectively.

  The fit parameters for eqs. (\ref{eq:Pol})-(\ref{eq:alpha}) are shown in 
 Table \ref{T1_lambda}  for different  fit conditions. The fits have been
performed for the data  \cite{Abe_D34}-\cite{Smith}.

In the fit \# 1 all  parameters are free.
 The use of  $p_{T}$-dependent phases  $x_{1}$ and 
$x_{2}$ in eqs.  (\ref{eq:x1})-(\ref{eq:x2}) improves the fit quality 
significantly ($\chi^{2}/N_{DOF}$ changed from 1.81 to 1.26). 

\begin{figure}[ht]
\centerline{\epsfig{file=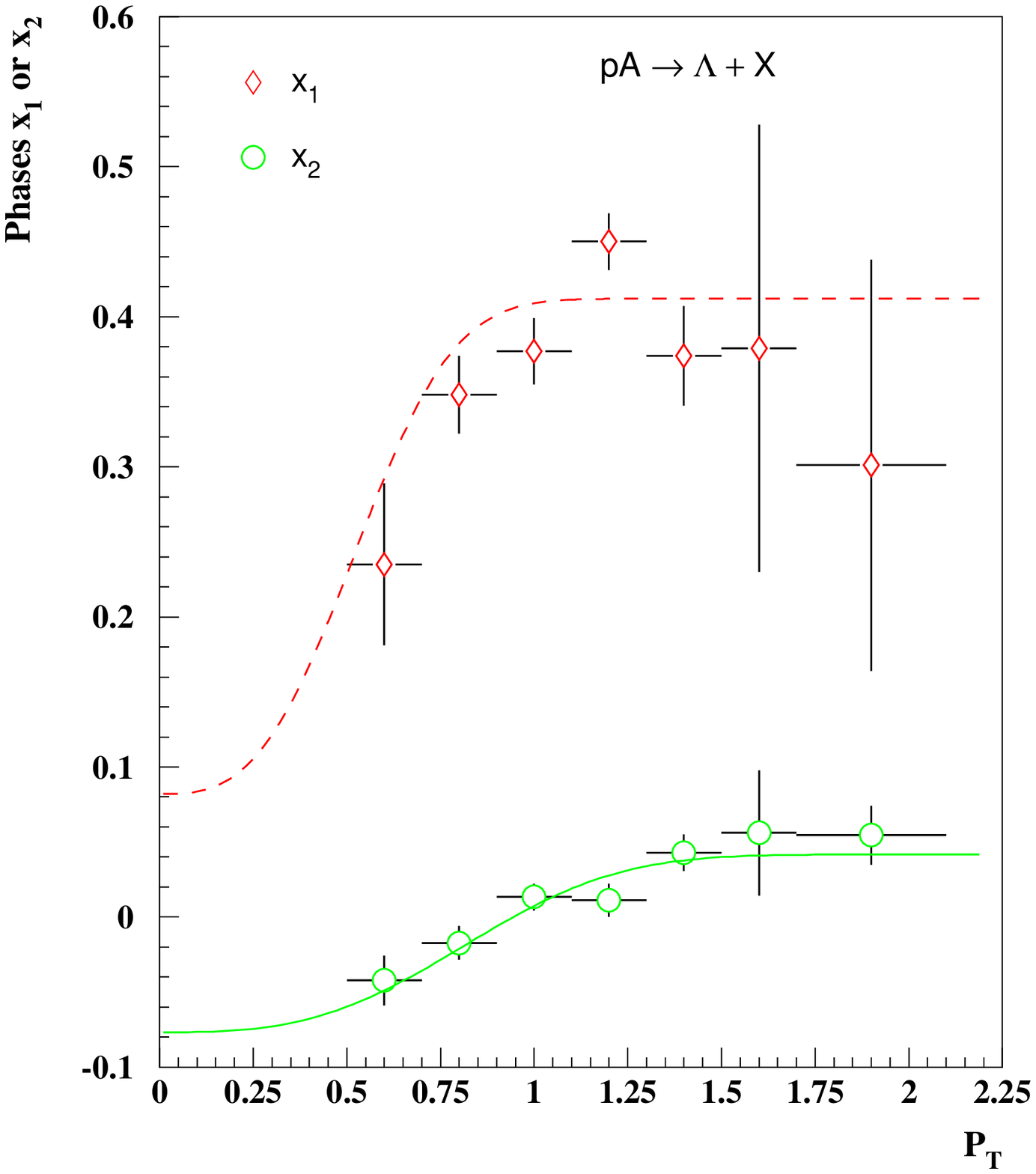,width=15cm}}
\caption {The dependence of $x_{1}$ and $x_{2}$ parameters on $p_{T}$
 for $\Lambda$ production in $pp(A)$ collisions. The curves correspond to
eq. (13) (dashed) and eq. (14) (dash-dotted), respectively.}
    \label{x1_x2_pt}
\end{figure}

\begin{figure}[ht]
\centerline{\epsfig{file=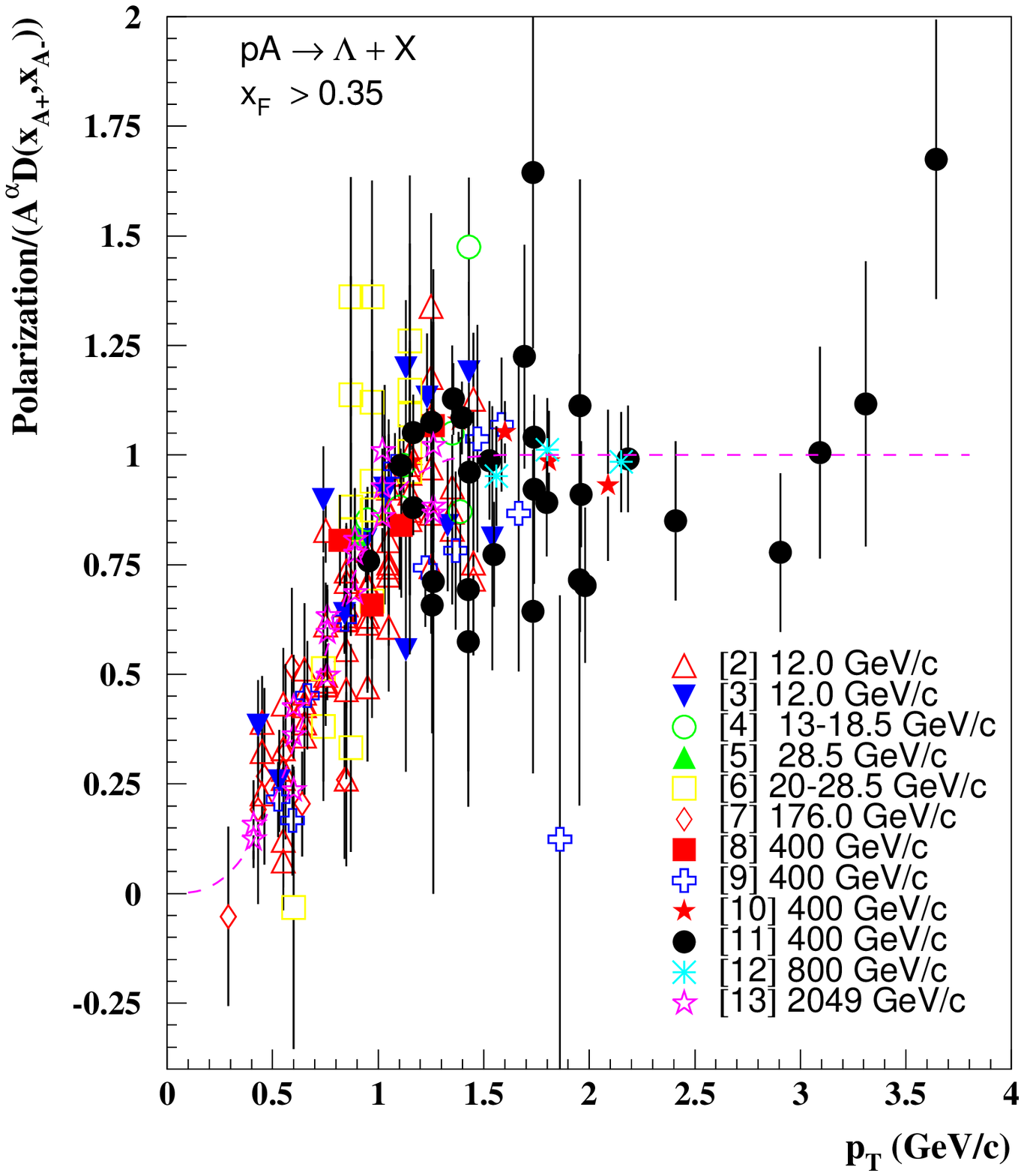,width=15cm}}
\caption {The ratio of polarization and $A^{\alpha}D(x_{A+},x_{A-})$,
 (eq. (18)) vs $p_{T}$ for $\Lambda$ production 
in $pp(A)$ collisions, and $x_{F} \ge 0.35$. 
 Parameters of eqs. (10)-(15) are presented in Table 1, fit \# 2.
 The  curve corresponds to the function $F(p_{T})$ in eq. (10). }
    \label{lam_pa_pt}
\end{figure}

 The dependence of phases $x_{1}$ and $x_{2}$ on $p_{T}$ is shown in 
Fig. \ref{x1_x2_pt}. As in the case of  polarization  dependence on $p_{T}$ 
(see Fig.  \ref{lam_pa_pt}) the phases $x_{1}$ and $x_{2}$ are described well
 by the terms $\propto [1- a\cdot exp(-b\cdot p_{T}^{3})]$ with  a plateau 
for $p_{T}$ above 1-2 GeV/c. The absolute value of $x_{2}$ is  small in 
accordance with the approximate validity of the feature (\ref{eq:Pasym})  
for $pA$ collisions.

 The values of $w_{1}=0.51\pm0.05$ 
and $\sigma=1.03\pm0.09$ for the fit \# 1  are close to the expected ones 
(0.5 and 1, respectively) with 
$\chi^{2}/N_{DOF}=1.27$. So, the fit \# 1 supports the choice of scaling
variables in the form $x_{A\pm} = (x_{R} \pm x_{F})/2$ and an approximate 
rotational asymmetry (\ref{eq:Pasym})  for the polarization in $pA$
 collisions, though the experimental data exist only in the forward
 hemisphere.  The parameter $\omega$  in 
the fit \# 1 is equal to $3.29 \pm 0.38$. The large value of $\omega$
 excludes the possibility of a linear dependence of polarization on
$x_{A\pm}$ or $x_{F}$ variables.

The fit \# 2 is made with four parameters
fixed ($\omega=3.045$, $\sigma=1$, $w_{1}=0.5$ and $\alpha_{0}=0$) that
simplifies  eq. (\ref{eq:Pol}) and allows one to
determine the other parameters with a better accuracy. 

 An interesting feature of the above fits is that at the first 
approximation the $\alpha$ is  proportional to $|x_{F}|$.
 The higher is a hyperon momentum  the more its polarization is 
attenuated by the interactions with a nuclear target.
 For a heavy nuclear target and a large $x_{F}$ we expect a
significant attenuation of the polarization in comparison with the $pp$ 
collisions case.
In particular, for $x_{F}=0.7$ we have $\alpha = -0.16|x_{F}|=0.109$ and the
polarization on a Lead target  is reduced by a factor $A^{-0.109} = 0.56$.
On a Beryllium target a corresponding factor is 0.79. A similar order of
magnitude for the polarization degradation in complex nuclei can be found
in \cite{Pondrom}, where it is also shown that the polarization degradation
is $p_{T}$-independent.
\begin{table}[htb]
\small
\caption{Fit parameters of eqs.~(10)-(15) for $\Lambda$ production
in $pp(A)$ collisions.
Different fit conditions are explained in the text. }
\begin{center}
\begin{tabular}{ccccccc}    \\[0.1cm] \hline 
Fit \#           &        1           &         2        
               \\[0.1cm] \hline  \\[0.0cm]
$c$              &-1.22$\pm$0.09   &
-1.22$\pm$0.05   &       \\ [0.1cm]

$\omega$         & 3.29$\pm$0.38   &
      $3.045$    &       \\ [0.1cm]

$\sigma$         & 1.03$\pm$0.09   &
      1.0        &          \\ [0.1cm]

$w_{1}$          & 0.51$\pm$0.05   &
      0.5        &         \\ [0.1cm]

$\eta_{0}$       & 0.404$\pm$0.034 &
 0.412$\pm$0.017 &    \\ [0.1cm]

$\eta_{1}$       & 0.29$\pm$0.14   &
 0.33$\pm$0.14   &     \\ [0.1cm]

$\delta$         & 4.6$\pm$1.7     &
 4.7$\pm$1.7     &       \\ [0.1cm]

$\xi_{0}$        & 0.077$\pm$0.025 &
 0.075$\pm$0.011 &    \\ [0.1cm]

$\xi_{1}$        & 0.218$\pm$0.061 &
 0.213$\pm$0.042 &    \\ [0.1cm]

$\nu$            & 1.26$\pm$0.23   &
 1.24$\pm$0.22   &     \\ [0.1cm]

$\kappa$         & 1.89$\pm$0.15   &
 1.88$\pm$0.13   &     \\ [0.1cm]

$\alpha_{0}$     & -0.01$\pm$0.03   &
       0.0       &      \\ [0.1cm]

$\alpha_{1}$     & -0.133$\pm$0.053&
-0.156$\pm$0.015 &   \\ [0.1cm]

$\chi^{2}/N_{DOF}$ & 325.7/256       &
   326.0/260           \\ [0.1cm] \hline
 \end{tabular}
\end{center}
\label{T1_lambda}
\end{table}
   The $p_{T}$ dependence of $\Pol_{\Lambda}$ is described by the
 function ${\FPT}(p_{T})$, which is sharply rising
for $p_{T} \le 1.2$ GeV/c and is practically constant for higher $p_{T}$.
 The experimental dependence of $\Pol_{\Lambda}$ on $p_{T}$
is illustrated in  Fig. \ref{lam_pa_pt} , where the ratio of $\Pol_{\Lambda}$
 and $ A^{\alpha}{\DRF}(x_{A+},x_{A-})$ is shown, and
\begin{equation}
  {\DRF}(x_{A+},x_{A-}) =  
{\GXA}(x_{A+} -x_{2}) - \sigma  {\GXA}(x_{A-} +x_{2}) .
\label{eq:DRF}
\end{equation}
 The ratio is assumed to be a function of $p_{T}$ only. An additional
cut $x_{F} \ge 0.35$ is used for the data shown in   Fig. \ref{lam_pa_pt}
 to exclude the points with large fractional errors of the ratio. The data in 
 Fig. \ref{lam_pa_pt} show the independence of the ratio on the beam energy,
 $x_{F}$ and the target type, and confirms the scaling behaviour
and  factorization of the $p_{T}$ and $x_{A\pm}$ dependencies, assumed in
(\ref{eq:Pol}).
New measurements are desirable for $p_{T} \ge 1.5$ GeV/c to clarify
the $\Pol_{\Lambda}$ and ${\FPT}(p_{T})$ behaviour at high $p_{T}$.

\begin{figure}[ht]
\centerline{\epsfig{file=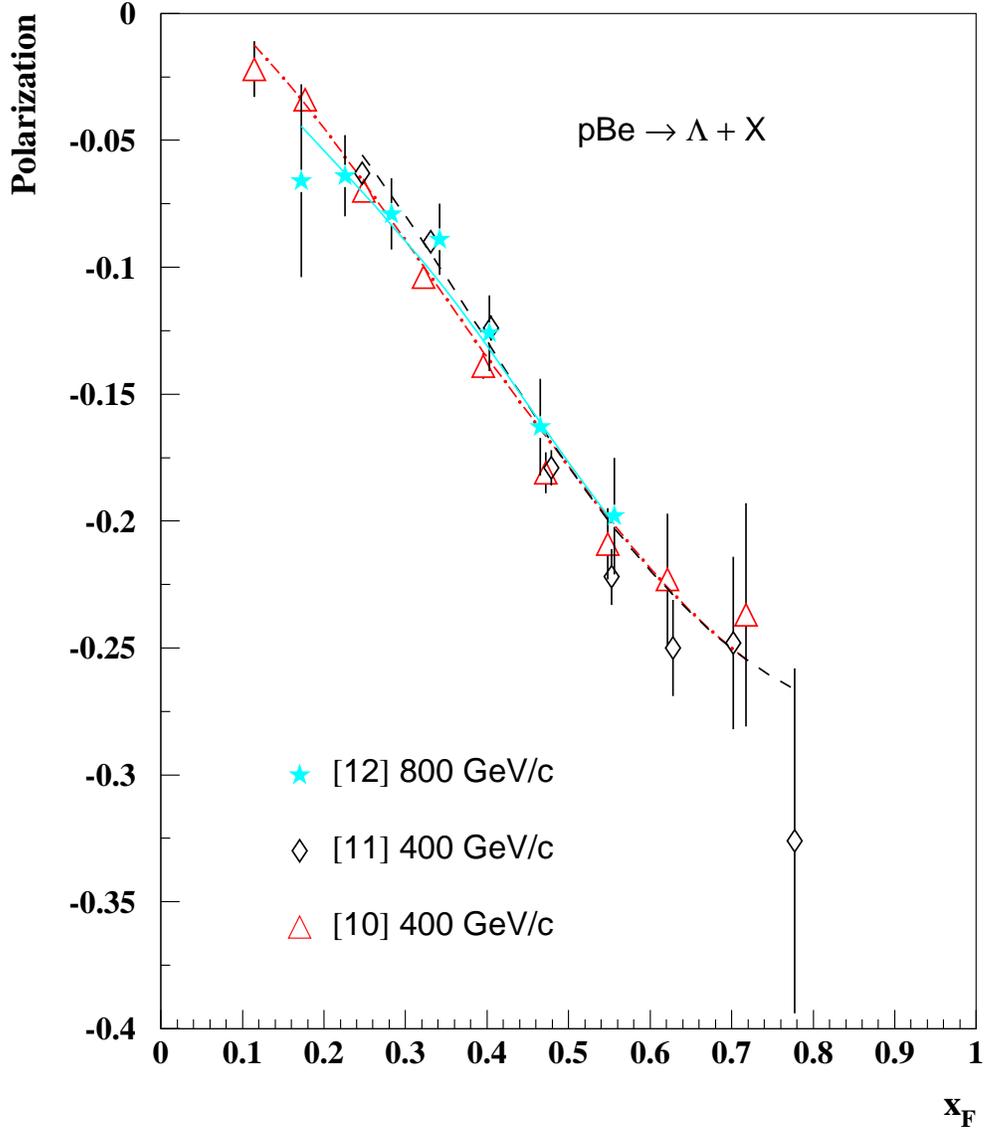,width=15cm}}
\caption {Polarization vs $x_{F}$ for $\Lambda$ production
in $pBe$ collisions. The fit parameters  of eq. (10)
are presented in Table 1, fit \# 2.  The fitting curves correspond to
data [12] (solid), [11] (dashed), and [10] (dash-dotted), respectively.}
    \label{lam_be_xf}
\end{figure}

The dependence of  $\Pol_{\Lambda}$ on $x_{F}$ is illustrated in Fig.
\ref{lam_be_xf}, where the most precise polarization data , measured in $pBe$
collisions at 400 and 800 GeV/c are shown 
\cite{{Heller_41},{Lundberg},{Ramberg}}. The fitting curves reproduce 
the $x_{F}$ dependence, as well as  small variations of it with $p_{T}$.

\begin{figure}[ht]
\centerline{\epsfig{file=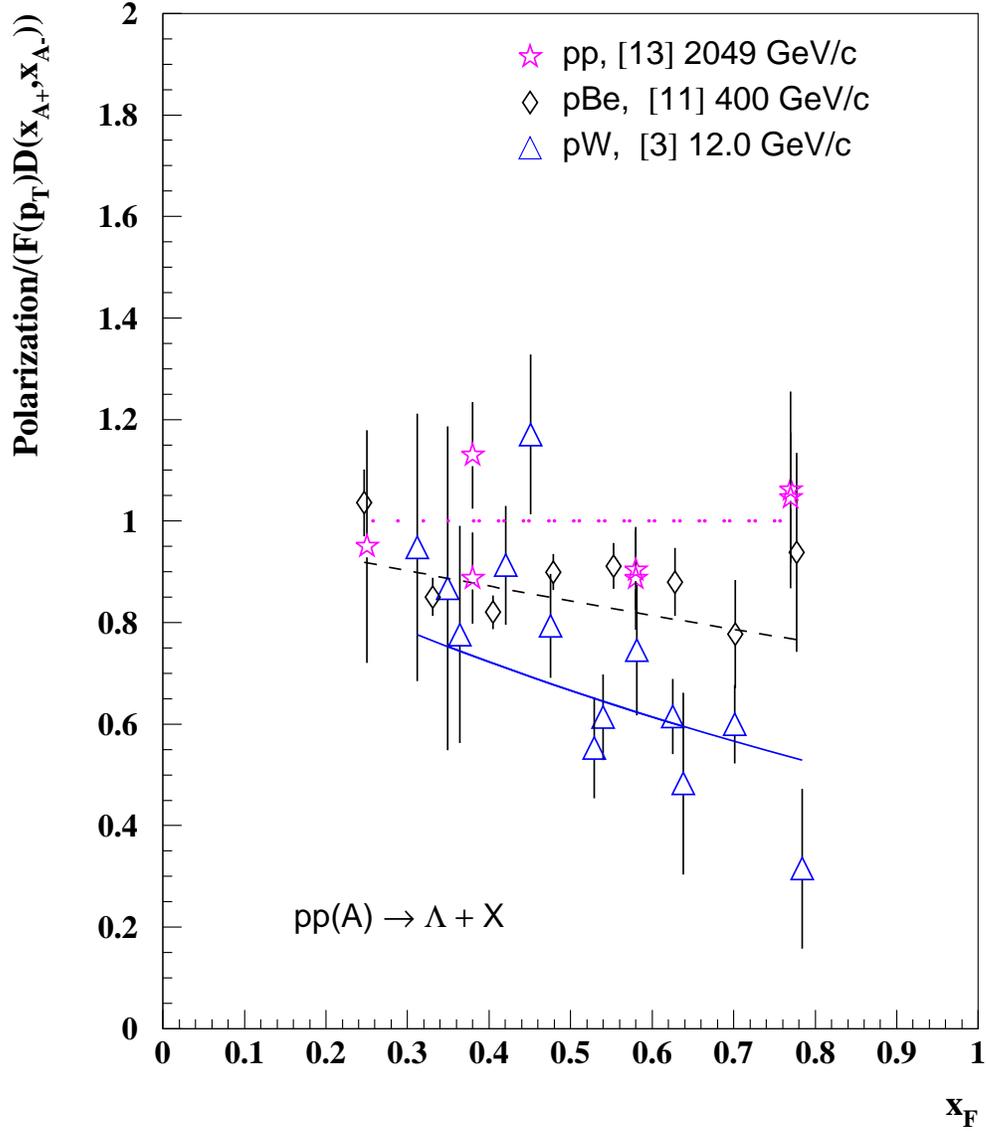,width=15cm}}
\caption {The ratio of polarization and $F(p_{T})D(x_{A+},x_{A-})$ 
(see eqs. (10)-(18))
 vs $x_{F}$ for $\Lambda$ production 
in $pp$ [13], $pBe$ [11] and $pW$ [3] collisions.
 Parameters of eqs. (10)-(15) are presented in Table 1, fit \# 2.
 The  curves correspond to the $pp$ (dotted), $pBe$ (dashed) and 
 $pW$ (solid) collisions, respectively.}
    \label{lam_adep_xf}
\end{figure}

The attenuation of the $\Lambda$ polarization on nuclear targets and 
the dependence of the attenuation  on $x_{F}$ are illustrated in  Fig. 
\ref{lam_adep_xf}. The measured polarization is divided by 
${\FPT}(p_{T}){\DRF}(x_{A+},x_{A-})$ and the corresponding ratio is plotted
 in  Fig. \ref{lam_adep_xf} vs $x_{F}$. The  data for 
the $pp$  \cite{Smith}, $pBe$ \cite{Lundberg}, and $pW$ \cite{Abe_PRL50}
collisions show a decrease of polarization on $Be$ and $W$ targets 
with $x_{F}$ rise in comparison with the proton-proton collisions case. 
The existing ${\Pol_{\Lambda}}$ data on medium and heavy nuclear targets are 
limited in terms of their accuracy and kinematic range. 
Additional measurements on medium and heavy nuclear targets for positive
and negative $x_{F}$ are desirable to confirm and clarify
the polarization attenuation effects, shown in Fig. \ref{lam_adep_xf}.
It will be interesting also to measure the $A$-dependence of the analyzing 
power ${\AN}$ and compare it with that of the hyperon polarization.

 The general agreement between the data and the fits \# 1 - \# 2 can be 
considered a good  one, taking into account  statistical and possible 
systematic errors of the data in different experiments.

 We may conclude
that $\Lambda$ polarization reveals a scaling behaviour, when it is
presented in a forward hemisphere as a function of two scaling variables
 $x_{A\pm}$ (or $x_{R}$ and $x_{F}$) and $p_{T}$.

\section{ Polarization of $\Sigma^{\pm ,0}$ and $\Xi^{0,-}$ hyperons 
in $pp$ and $pA$ collisions}

In this section the polarization of $\Sigma^{\pm ,0}$ and $\Xi^{0,-}$ hyperons
in $pp$ and $pA$ collisions is analyzed. The corresponding fit parameters
are presented in Tables \ref{T2_sigma} and \ref{T2_Xi}. The fits have been
 performed using eqs. (\ref{eq:Pol})-(\ref{eq:alpha}), and eq. 
\begin{equation}
  {\FPT}(p_{T})  = \cases
 { p_{T}/p_{T}^{0}    , & if $p_{T} \le p_{T}^{0}$;\cr
   1    , & otherwise;\cr} \quad   \label{eq:F2}
\end{equation}
 for the function ${\FPT}(p_{T})$, which is found to be more appropriate
for an approximation of the polarization dependence on $p_{T}$. 
For this section and all the following  the parameters $\alpha_{0}$ and 
$w_{1}$ are fixed ( $\alpha_{0}=0$ and $w_{1}=0.5$).
\subsection{The $\Sigma^{+}$ hyperon polarization}
The polarization of $\Sigma^{+}$ hyperons ($\Pol_{\Sigma+}$)
in $pA$ collisions has been measured on a $Cu$ target at 800 GeV/c 
\cite{Morelos} and on a $Cu$ and a $Be$ targets at 400 GeV
 \cite{Wilk,Ankenb,Wilk_46}.  A fit \# 1 is made  with
some parameters fixed due to a limited data statistics.
 The parameter $\omega=1.9\pm 1.2$
 is compatible with the value of $\omega$ for the case of $\Lambda$ 
polarization discussed in the  previous section. The fit \# 2 is done
with the value of $\omega=3.045$, which follows from eq. (\ref{eq:omega}).

\begin{figure}[ht]
\centerline{\epsfig{file=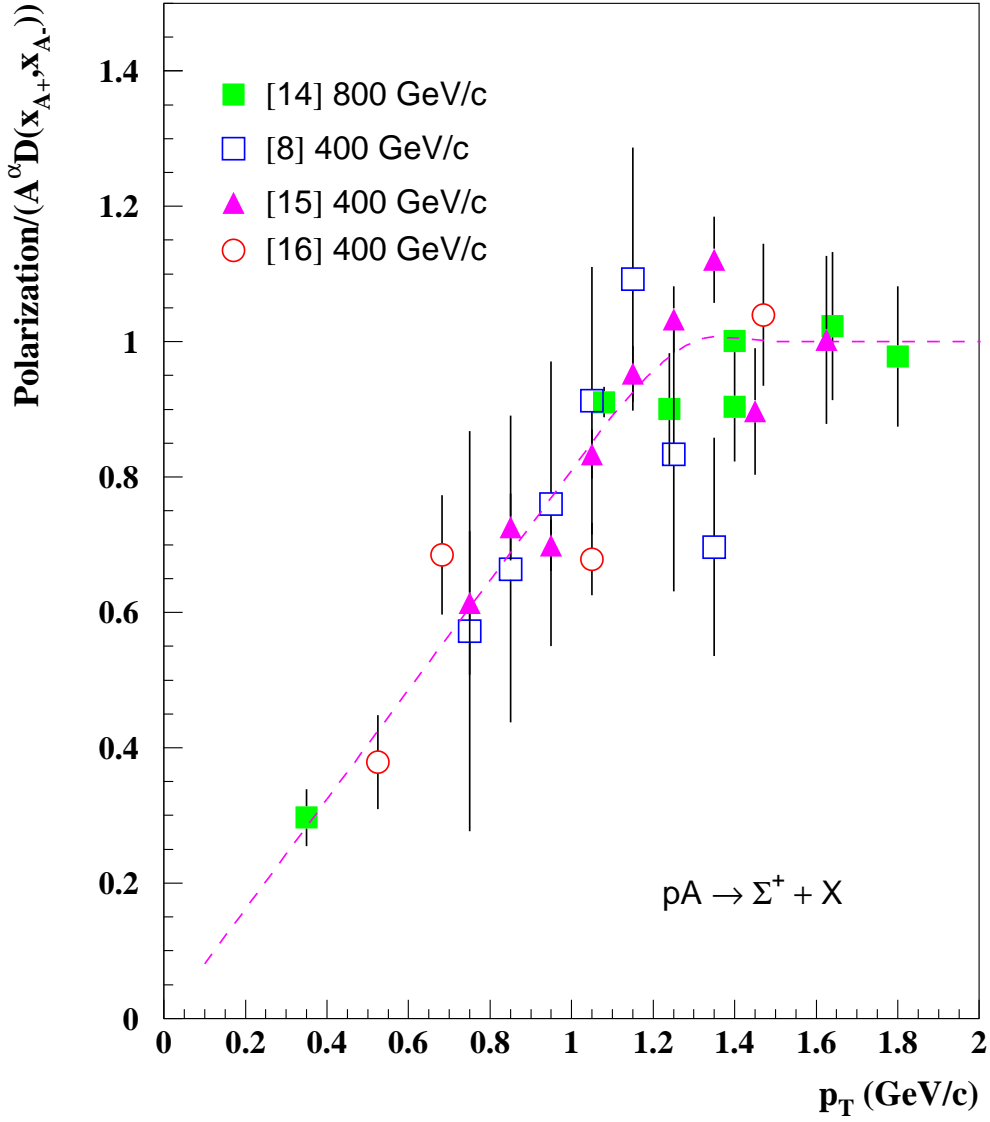,width=15cm}}
\caption {The ratio of polarization and $A^{\alpha}D(x_{A+},x_{A-}$),
 (eq. (18)) vs $p_{T}$ for $\Sigma^{+}$ production in $pA$ collisions. 
 Parameters of eqs. (11)-(19) are presented in Table 2, 
fit \# 2. The  curve corresponds to the function $F(p_{T})$ in eq. (19). }
    \label{sigp_pa_pt}
\end{figure}

The dependence of $\Pol_{\Sigma+}$ on $p_{T}$ is illustrated 
in  Fig. \ref{sigp_pa_pt}, where a ratio of $\Pol_{\Sigma+}$ and
$A^{\alpha}{\DRF}(x_{A+},x_{A-})$ is plotted.
The  curve corresponds to the function ${\FPT}(p_{T})$ in eq. (\ref{eq:F2})
with $p_{T}^{0} \approx 1.2$ GeV/c (fit \# 2).  The function ${\FPT}(p_{T})$
 is well approximated  by a linear dependence for $p_{T}\le 1.2$  GeV/c 
and a plateau for higher  $p_{T}$ (see Fig. \ref{sigp_pa_pt}), but additional
 measurements are desirable for higher $p_{T}$ and different $x_{F}$.

It was stated in \cite{Morelos} that
 the $\Sigma^{+}$ polarization at 800 GeV/c decreases as a function
 of $p_{T}$ at fixed $x_{F}$. The results of the fit \# 2 indicate that
such unusual $p_{T}$ dependence is probably due to
the $p_{T}$ dependence of the  parameter
 $x_{2}$ in eq. (\ref{eq:x2}).  As we can see from eq. (\ref{eq:Pol1}),
the polarization (for $\sigma=1$) is proportional to 
$\sin [\omega(x_{F}w_{2} - x_{2}) ] \approx \sin [3(x_{F}/2 - x_{2}) ]$.
Since $x_{2} \approx (1-Z/A)(0.34 - 0.17 e^{ -0.82 p_{T}^{3} })$ and 
$(1-Z/A) \approx 0.4$, the value of $x_{2}$ is about 0.09 for 
$p_{T} \le 0.5$ GeV/c, it
starts to grow fast for  $0.5 \le p_{T} \le 1.5$ GeV/c, and  has a 
plateau $x_{2} \approx 0.19$ for higher  $p_{T}$ values. The value of $x_{F}$
 in \cite{Morelos} is about 0.46, so ${\Pol_{\Sigma+}} \propto
 \sin [3(0.23 -x_{2}) ]$ and decreases with $p_{T}$ rise for 
$p_{T} \ge 0.7$ GeV/c. This decrease of ${\DRF}(x_{A+},x_{A-})$ is 
not compensated by a corresponding increase of the ${\FPT}(p_{T})$, that leads
to the observed polarization decrease with $p_{T}$ rise.  If we take 
$x_{F} \ge 0.65$, such effect as a decrease of polarization with $p_{T}$ 
increasing  is not expected. The results of three 
other $\Sigma^{+}$ polarization  measurements, which have typical
 $x_{F} \ge 0.52$, do not reveal a decrease  of $\Pol_{\Sigma+}$ with 
$p_{T}$ rise \cite{Wilk,Ankenb,Wilk_46}.
\begin{table}[htb]
\small
\caption{Fit parameters of eqs.~(10)-(19) for $\Sigma^{\pm}$
production in $pp(A)$ collisions.}
\begin{center}
\begin{tabular}{cccccccc}  \\[0.1cm]  \hline

Fit \#           &   1                &  2              &  3            & 
         4               \\[0.1cm] \hline  \\[0.1cm]

$H$              &     $\Sigma^+$     &  $\Sigma^+$     &   $\Sigma^-$  &
   $\Sigma^-$                 \\ [0.1cm]

$c$              &    4.4$\pm$1.0     &  4.0$\pm$0.9     &4.2$\pm$2.5   &
 4.2$\pm$4.2                \\ [0.1cm]

$\omega$         &    1.9$\pm$1.2     &     $3.045$      &  6.1$\pm$4.0 &
 6.090                      \\ [0.1cm]

$\sigma$        &       1.0          &      1.0         &    1.0       &
       1.0                    \\ [0.1cm]

$p_{T}^{0}$      &    1.21$\pm$0.07   &  1.24$\pm$0.06   &    0.66      &
 0.66                   \\ [0.1cm]

$\eta_{0}$       & -0.10$\pm$0.31     &  0.107$\pm$0.059 &    0.25       &
 0.25$\pm$0.20           \\ [0.1cm]

$\eta_{1}$       & -0.56$\pm$0.79     &  -0.22$\pm$0.62  &    0.0       &
 0.0                  \\ [0.1cm]

$\delta$         &  4.7$\pm$7.5       &    5$\pm$19      &    0.0       &
  0.0                \\ [0.1cm]

$\xi_{0}$        &  0.337$\pm$ 0.015  & 0.341$\pm$ 0.013 &0.307$\pm$0.058 &
 0.31$\pm$0.11              \\ [0.1cm]

$\xi_{1}$        &  0.172$\pm$0.054   &  0.198$\pm$0.044 &   0.0         &
 0.0                   \\ [0.1cm]

$\nu$       &  0.82$\pm$0.28     &  0.98$\pm$0.20   &   0.0         &
 0.0                   \\ [0.1cm]

$\alpha_{1}$     & -0.26$\pm$0.05     & -0.24$\pm$0.04   &  -0.26        &
-0.26               \\ [0.1cm]

$\chi^{2}/N_{DOF}$ & 33.0/17          &   33.6/18        &  0.06/3        &
 0.06/3                        \\ [0.1cm] \hline

 \end{tabular}
\end{center}
\label{T2_sigma}
\end{table}
\begin{table}[htb]
\small
\caption{Fit parameters of eqs.~(10)-(19) for 
 $\Xi^{-,0}$ production in $pp(A)$ collisions.}
\begin{center}
\begin{tabular}{cccccccc}  \\[0.1cm]  \hline

Fit \#           &   1                & 
 2               &  3              &          4  \\[0.1cm] \hline  \\[0.1cm]

$H$              &   $\Xi^-$  &
   $\Xi^-$       &     $\Xi^0$     &     $\Xi^0$                  \\ [0.1cm]

$c$              &  -0.81$\pm$0.12  &
-0.81$\pm$0.12   & -1.37$\pm$0.50    & -1.37$\pm$0.61           \\ [0.1cm]

$\omega$         &    5.95$\pm$0.74  &
   6.090          &  $6.10\pm0.92$    &   6.090            \\ [0.1cm]

$\sigma$        &       1.0        &
       1.0       &       1.0          &      1.0                 \\ [0.1cm]

$p_{T}^{0}$      &    0.4$\pm$0.2    &
 0.4$\pm$0.3     &    0.5            &      0.5                 \\ [0.1cm]

$\eta_{0}$       & 0.29$\pm$0.11      &
0.284$\pm$0.073  & 0.336$\pm$0.023    & 0.336$\pm$0.036          \\ [0.1cm]

$\eta_{1}$       & -0.030$\pm$0.082   &
-0.033$\pm$0.060 &       0.0          &       0.0                 \\ [0.1cm]

$\delta$         &  2 $\pm$12         &
  2.5$\pm$9.7    &       0.0          &       0.0                 \\ [0.1cm]

$\xi_{0}$        &  0.00$\pm$0.22     &
 0.02$\pm$0.13   &      -0.2          & -0.20$\pm$0.15          \\ [0.1cm]

$\xi_{1}$        & 0.51$\pm$0.12      &
 0.50$\pm$0.10   &     0.0            &        0.0                \\ [0.1cm]

$\nu$       &  3.0$\pm$1.9       &
 2.9$\pm$1.5     &      0.0           &       0.0                 \\ [0.1cm]

$\alpha_{1}$     &  0.03$\pm$0.14     &
  0.04$\pm$0.11  &     -0.12          &     -0.12                 \\ [0.1cm]

$\chi^{2}/N_{DOF}$ & 71.0/46          &
 71.0/47           &  5.7/13          &     5.7/13    \\ [0.1cm] \hline
 \end{tabular}
\end{center}
\label{T2_Xi}
\end{table}

It was also stated in \cite{Morelos} that an energy dependence of
 $\Pol_{\Sigma+}$ is observed by comparing the results at 800 GeV/c 
($Cu$ target) \cite{Morelos},  at 400 GeV/c ($Be$ target) \cite{Wilk} and 
400 GeV/c ($Cu$ target) \cite{Ankenb}.
 The observed difference   is  really due to the different targets 
($Cu$ vs $Be$) and slightly different $x_{F}$ values used for this comparison.
In the case of comparison of the data on $Be$ and $Cu$ targets a strong
$A$-dependence is the reason of a higher polarization on the $Be$ target,
since $\alpha = -0.26|x_{F}|$ (see fit \# 1). For the data on $Cu$ target
\cite{Ankenb} the corresponding $x_{F} \approx 0.52$ and 
${\Pol_{\Sigma+}}=0.168 \pm 0.017$, while the 800 GeV/c
data are measured at $x_{F} \approx 0.46$ and 
${\Pol}_{\Sigma+}=0.124 \pm 0.001$. The data \cite{Wilk}
show that the ${\Pol_{\Sigma+}}$ increases by 0.05 or more, when $x_{F}$ is
 increased from 0.47 to 0.52. So, the expected $\Pol_{\Sigma+}$ for 800 GeV/c
 and $x_{F} = 0.52$ is about $0.124 + 0.05 = 0.174$, which is compatible with
 the measured value ${\Pol_{\Sigma+}}=0.168 \pm 0.017$ at 400 GeV/c.

  The energy independence of the polarization  is confirmed by
 Fig. \ref{sigp_pa_pt}, where all the data points at two beam energies
 and two different targets are approximated well by a single function 
${\FPT}(p_{T})$ and are compatible with the scaling law (\ref{eq:Pol}), 
described  using $p_{T}$ and two  scaling variables $x_{A+}$ and $x_{A-}$.

The data fit indicates also that $\Sigma^{+}$ polarization 
decreases on a nuclear target ($\propto A^{-0.26|x_{F}|}$). The polarization 
attenuation is more significant at high $|x_{F}|$ values, similar to that of 
for  the $\Lambda$ hyperons.

 We may conclude that the use of two scaling variables ($x_{A+}$ and $x_{A-}$)
in the form of eq. (\ref{eq:Pol}) instead of one ($x_{F}$) resolves 
the problem of energy dependence of  the $\Sigma^{+}$ polarization and
 the problem of its unusual $p_{T}$ dependence
(see details in \cite{Morelos}), and presents the existing 
data in the energy independent way.

\subsection{The $\Sigma^{-}$ hyperon polarization}
  The polarization of $\Sigma^{-}$ ($\Pol_{\Sigma-}$) has been measured 
at 400 GeV/c in $pCu$ \cite{Wah} and $pBe$ \cite{Deck} collisions. Since 
just a few points in $x_{F}$ and $p_{T}$ has been measured, some of the 
fit \# 3  parameters are fixed, see
 Table \ref{T2_sigma}. In addition, for the fit \# 4 the $\omega$ parameter was
 fixed: $\omega=6.090$. According to the fit
 the    $\Pol_{\Sigma-}$  has a maximum near
$x_{F}\approx 0.67$ with corresponding $\omega \approx 6$. 

\subsection{The $\Sigma^{0}$ hyperon polarization}
  The polarization of $\Sigma^{0}$ ($\Pol_{\Sigma0}$) produced  in $pBe$ 
has been measured at 28.5 GeV/c \cite{Dukes} and at 18.5 GeV/c \cite{Bonner1}.
Only two data points are available from these two experiments.
At 28.5 GeV/c the value of $\Pol_{\Sigma0}$ is 
$0.28 \pm 0.13$ ($x_{F}=0.6$, $p_{T}=1.01$ GeV/c ), and at 18.5 GeV/c it is
$0.23 \pm 0.13$ ($0 < x_{F}<0.75$, $0.5 < p_{T} < 2$ GeV/c), which is 
consistent with the $\Sigma^{\pm}$ polarization in the same kinematic area.
 
\subsection{The $\Xi^{-}$ hyperon polarization}
\begin{figure}[ht]
\centerline{\epsfig{file=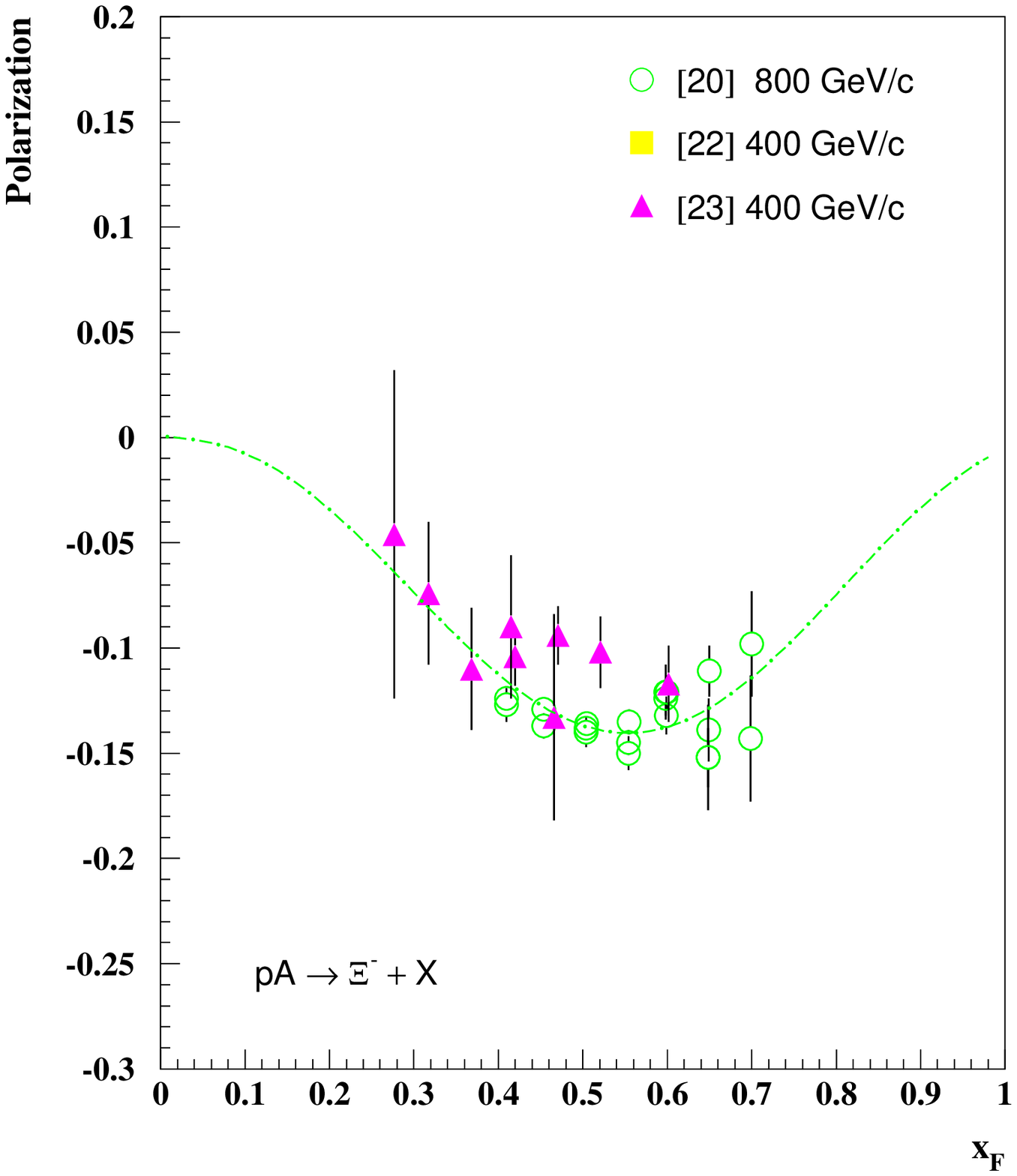,width=15cm}}
\caption {Polarization vs $x_{F}$ for $\Xi^{-}$ production 
in $pA$ collisions. The fit parameters of eqs. (10)-(19) are
presented in Table 3. The  curve correspond to the fit \# 2 for
 $p_{T}=1.5$ GeV/c.}
    \label{xim_pa_xf}
\end{figure}

 The polarization of $\Xi^{-}$ ($\Pol_{\Xi-}$) has been measured at 800 GeV/c
in $pBe$ \cite{Duryea,Ho} collisions, and at 400 GeV/c in $pCu$ \cite{Trost}
and $pBe$ \cite{Rameika} collisions. 
The fit \# 1 is made with some parameters fixed. 
The value of parameter $\omega$  is $5.95 \pm 0.74$.
In the fit \# 2  the  $\omega=6.090$ is used 
and the corresponding curve is shown in  Fig. \ref{xim_pa_xf} for
 400 GeV/c $pBe$ collisions and $p_{T}=1.5$ GeV/c. The data shown in 
Fig.  \ref{xim_pa_xf} have an additional cut $0.8 < p_{T} < 1.5$ GeV/c
to reduce the smearing of the data points due to the $p_{T}$ dependence.

The fit \# 2 and  Fig. \ref{xim_pa_xf} data indicate a local maximum 
in the absolute value of polarization at $x_{F} \approx 0.5$.

 The $A$-dependence of the $\Pol_{\Xi-}$ is not significant 
($\alpha_{1} \approx 0.04 \pm 0.11$).

\subsection{The $\Xi^{0}$ hyperon polarization}
\begin{figure}[ht]
\centerline{\epsfig{file=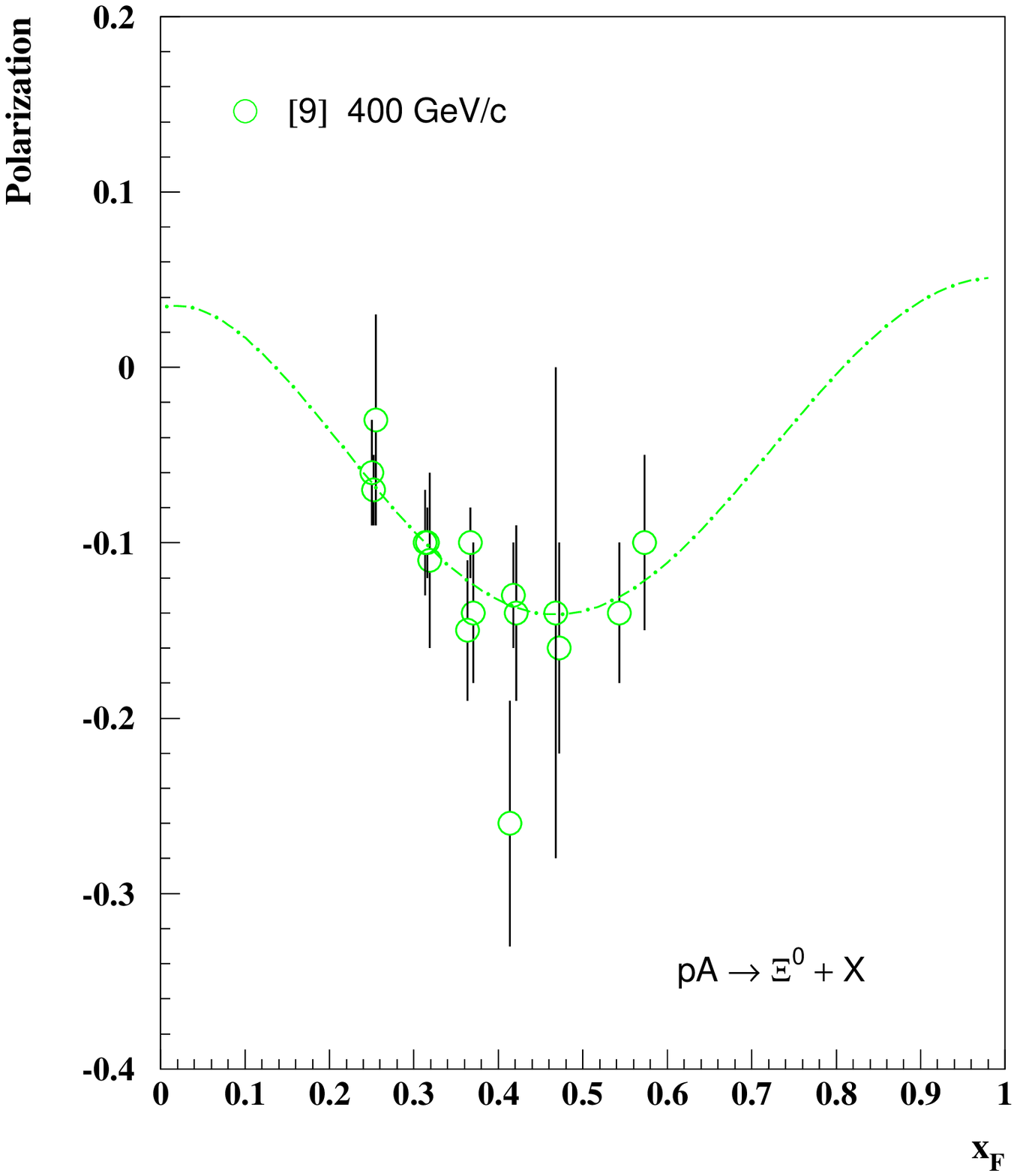,width=15cm}}
\caption {Polarization vs $x_{F}$ for $\Xi^{0}$ production 
in $pA$ collisions. The fit parameters of eqs. (10)-(19)  are
presented in Table 3. The  curve correspond to the fit \# 4 for
 $p_{T}=1.5$ GeV/c.}
    \label{xi0_pa_xf}
\end{figure}

 The polarization of $\Xi^{0}$ ($\Pol_{\Xi0}$) has been measured at 400 GeV/c
in $pBe$ \cite{Heller_51} collisions.  The data are presented in 
 Fig. \ref{xi0_pa_xf} vs $x_{F}$. Due to a small number of experimental
points some of the fit parameters were fixed near the values which give
the best $\chi^{2}$. The fit \# 3 is done with the $\omega$ free and
the fit \# 4 - with $\omega=6.090$.

 The fit \# 4 and  Fig.  \ref{xi0_pa_xf} data indicate a local maximum in 
the absolute value of polarization near $x_{F} \approx 0.46$.

The fits \# 2 and \# 4 indicate a possible decrease in the absolute value of 
$\Pol_{\Xi0}$ and $\Pol_{\Xi-}$ for $x_{F} \ge 0.6-0.8$ due to a high value
of $\omega \approx 6$. This feature makes these reactions different from
the $\Lambda$ production in $pA$ collisions, where we see a steady increase
of polarization with $x_{F}$ rise.

  An interesting observation follows from the analysis of the scaling 
properties of  hyperon ($\Lambda$, $\Sigma^{\pm}$, $\Xi^{-}$ and $\Xi^{0}$) 
polarization in $pp(A)$ collisions -  the parameter $\omega$ 
(``polarization oscillation frequency'') in eqs. 
(\ref{eq:Pol})-(\ref{eq:F2}), which describes the rate of change of
 the polarization with $x_{A\pm}$ increase, is related with the number
 of sea quarks ($N_{SEA}$), picked up from the sea during a hyperon
 production: $\omega \approx 3 N_{SEA}$.

%
\section{Polarization of $\Lambda$, $\Xi^{-}$ and $\bar{\Lambda}$ 
 hyperons produced  in $K^{-}p$  collisions}
%
 In this section we analyze the hyperon polarization data in $K^{-}p$ 
collisions. The corresponding fit parameters are
presented in  Table \ref{T5_km}. Since the initial state ($K^{-}p$) 
is not symmetric vs the rotation transformation and there are no
measurements on nuclear targets, some modification of eqs. 
(\ref{eq:Pol})-(\ref{eq:F2}) is introduced. In particular, the parameters
$\alpha_{0}$, $\alpha_{1}$ are fixed at zero values, $w_{1}=0.5$, and 
 the factor $(1-Z/A)$ in eq. (\ref{eq:x2}) is omitted:
\begin{equation}
  x_{2} = \xi_{0} - \xi_{1} e^{ -\varepsilon p_{T}^{3} }.
\label{eq:x2N}
\end{equation}
The function ${\FPT(p_{T})}$ is given by eq. (\ref{eq:F2}).
The $\sigma$ parameter in (\ref{eq:Pol}) could be different from unity
 due to an asymmetric initial state.
\subsection{The $\Lambda$ polarization in $K^{-}p$ collisions}
  The polarization of $\Lambda$ ($\Pol_{\Lambda}$) hyperons in $K^{-}p$ 
collisions has been measured at 176 GeV/c \cite{Gourlay},
at 32 GeV/c  \cite{Faccini}, at 14.3 GeV/c \cite{Borg,Abramowicz},
and at 10 and 16 GeV/c \cite{Grassler}. The data with lower beam momenta
are not used for this analysis due to the energy dependence of the polarization
below 9 GeV/c at fixed negative $x_{F}$ 
\cite{Faccini,Borg,Abramowicz,Baubillier}.
The interesting feature of these $K^{-}$ beam data is that they include
both, the beam and the target fragmentation regions. That feature allows us to
estimate the $\sigma$ parameter in  eq. (\ref{eq:Pol}) and check that
the scaling behaviour is valid for both hemispheres.

Only the data with $|x_{F}| \le 0.90$ are used for this analysis to avoid
a large contribution of exclusive channels near the kinematic limits.

 The data fits (\# 1 and \# 2) have been performed  with some
 parameters  fixed due to a limited statistics. 
In the fit \# 1 the parameter $\omega$ is
 free, and in the fit \# 2 its value is  fixed at $\omega=3.58$. Both fits have
 $\chi^{2}/N_{DOF}$ about 1.6 and indicate the
 maximum near $x_{F}=0.7$ and almost 
$x_{F}$-independent polarization for the negative $x_{F}$ region.
 The $\sigma$ parameter is $negative$ in  contrast to the $pp$ data.

The $p_{T}$ dependence of the $\Pol_{\Lambda}$ is compatible with a linear
 rise of the  ${\FPT}(p_{T})$ (see eq. (\ref{eq:F2})) for $p_{T} \le 0.85$
GeV/c.
\begin{table}[htb]
\small
\caption{Fit parameters of eqs.~(10)-(20) for $\Lambda$ and
 $\Xi^{-}$   production in $K^{-}p$ collisions.}
\begin{center}
\begin{tabular}{cccccccc}  \\[0.1cm]  \hline
Fit \#           &       1          &         2        &           3     & 
         4                       \\[0.1cm] \hline  \\[0.1cm]

$H$              & $\Lambda$        & $\Lambda$        & $\Xi^{-}$         &
  $\Xi^{-}$            \\ [0.1cm]

$c$              &  4.18 $\pm$0.47  & 4.25$\pm$0.37    & 6.0$\pm$2.8       &
  6.8$\pm$2.7          \\ [0.1cm]

$\omega$         &  3.53 $\pm$0.22  &  3.58            &3.04$\pm$0.88      &
  3.58                         \\ [0.1cm]

$\sigma$        & -0.77 $\pm$0.12  & -0.77 $\pm$0.11  &-0.90$\pm$0.67     &
-1.09$\pm$0.51                  \\ [0.1cm]

$p_{T}^{0}$     &        1.0        &        1.0       &     1.0           &
       1.0                   \\ [0.1cm]

$\eta_{0}$       & 0.076$\pm$0.019  &  0.079$\pm$0.016&0.127$\pm$0.049     &
 0.142$\pm$0.038        \\ [0.1cm]

$\eta_{1}$       &  0.81 $\pm$0.21  &  0.80 $\pm$0.20 &-0.01 $\pm$0.13     &
-0.046$\pm$0.065                   \\ [0.1cm]

$\delta$         &  26              &      26          &       26         &
  26                         \\ [0.1cm]

$\xi_{0}$        &  0.191$\pm$0.063 & 0.19$\pm$0.06    & 0.06$\pm$0.16     &
 0.09$\pm$0.08       \\ [0.1cm]

$\xi_{1}$        &  0.18 $\pm$0.13  & 0.19 $\pm$0.11   &-0.42$\pm$0.22     &
 -0.34$\pm$0.12            \\ [0.1cm]

$\nu$       &        7         &         7        &         27        &
    27                      \\ [0.1cm]

$\chi^{2}/N_{DOF}$ &    88.8/55     &      88.9/56     & 12.9/11     &
      13.3/12             \\ [0.1cm]  \hline

 \end{tabular}
\end{center}
\label{T5_km}
\end{table}
\subsection{The $\Xi^{-}$ polarization  in $K^{-}p$ collisions}
  The polarization of $\Xi^{-}$ ($\Pol_{\Xi -}$) hyperons in $K^{-}p$ 
has been studied at 5 GeV/c \cite{Bensinger} and at 4.2 GeV/c
 \cite{Ganguli}. Since the polarization
in \cite{Bensinger} is  integrated over one of the variables
($x_{F}$ or $p_{T}$), we have to assign  mean values for these
integrated variables. As  estimates of these mean values
we take for the analysis $x_{F} = 0.6$, and $p_{T} = 0.3$ GeV/c,
respectively.
Only data with $ |x_{F}| \le 0.85$ are used for the analysis to exclude
the resonance region, in particular in the extreme forward region 
($x_{F} \approx 1$) which is dominated by the baryon exchange process 
$K^{-}p \rightarrow \Lambda \pi^{0}$. The fit \# 3 is made with 
the $\omega$ parameter free, and the fit \# 4 is made for $\omega=3.58$.
As in the case of the $\Lambda$ polarization the $\sigma$ parameter is
negative, but with a much larger uncertainty.

Additional measurements are desirable for $p_{T} \ge 0.5$ GeV/c where
polarization could reach high values.
Both, the $x_{F}$ and the $p_{T}$ dependencies of $\Pol_{\Xi -}$ are
similar to the case of the $\Lambda$ polarization in $K^{-}p$ collisions.
\subsection{The $\bar{\Lambda}$ polarization  in $K^{-}p$ collisions}
The $\bar{\Lambda}$ polarization  in $K^{-}p$ collisions has been measured 
at 32 GeV/c \cite{Faccini}. Since just two $x_{F}$ points are available,
only the magnitude of polarization is estimated which is about 0.4.
\section{Hyperon polarization in $K^{+}p$  collisions}
%
  The hyperon polarization data for $K^{+}p$ collisions have been analyzed
using eqs. (\ref{eq:Pol})- (\ref{eq:x2N}). Only the data with 
 $|x_{F}| \le 0.74$ are used for the analysis. The fit parameters are presented
in  Table \ref{Tab_kp}. 
\subsection{The $\Lambda$ polarization in $K^{+}p$ collisions}
     The $\Lambda$ polarization in $K^{+}p$ collisions has been measured
at 8.2 and 16 GeV/c \cite{Chliapnikov}, at 13 GeV/c \cite{Barletta}, and
 at 32 GeV/c \cite{Faccini}. The fit \# 1 is made with the $\omega$ parameter
 free, and the fit \# 2 - with the $\omega=3.58$. 
 The polarization does not depend
on energy and has a maximum near $x_{F}=0.35$.
\subsection{The $\bar{\Lambda}$ polarization in $K^{+}p$ collisions}
   The $\bar{\Lambda}$ polarization in $K^{+}p$ collisions has been measured
at 32 and 70 GeV/c \cite{Ajinenko}, at 8.2 and 16 GeV/c \cite{Chliapnikov},
at 32 GeV/c \cite{Faccini}, and at 13 GeV/c \cite{Barletta}.
The data \cite{Ajinenko} for 32 and 70 GeV/c are combined for this analysis
and an average momentum 50 GeV/c is assigned to it.
 The fit \# 3 and the fit \# 4 
have been performed with $\omega$ parameter free and  fixed ($\omega=5.55$), 
respectively. The fits result in a negative $\sigma \approx -1$,
similar to the   $K^{-}p \rightarrow \Lambda$ case. The fits indicate the
existence of significant phases $x_{1}$ and $x_{2}$, which depend on $p_{T}$.
The hyperon polarization observed in this reaction is higher than in any 
other one.
\begin{table}[htb]
\small
\caption{Fit parameters of eqs.~(10)-(20) for $\Lambda$,
and $\bar{\Lambda}$ production in $K^{+}p$ collisions.}
\begin{center}
\begin{tabular}{cccccc}   \\[0.1cm]  \hline
Fit \#           &       1          &         2        &           3       & 
         4             \\[0.1cm] \hline  \\[0.1cm]

$H$              & $\Lambda$        & $\Lambda$        & $\bar{\Lambda}$   &
 $\bar{\Lambda}$    \\ [0.1cm]

$c$              & -3.0  $\pm$1.9   & -2.4$\pm$1.3      & 3.4$\pm$1.1       &
 3.9$\pm$1.0        \\ [0.1cm]

$\omega$         &  4.4  $\pm$1.5   &   3.58           &4.4$\pm$1.9        &
   5.55                      \\ [0.1cm]

$\sigma$        &  0.30 $\pm$0.18  &  0.36 $\pm$0.28  &-1.24$\pm$0.39    &
-1.40$\pm$0.28               \\ [0.1cm]

$p_{T}^{0}$     &        0.4       &        0.4       &     0.4          &
       0.4           \\ [0.1cm]

$\eta_{0}$       &-0.19 $\pm$ 0.13  & -0.21$\pm$0.13   &0.12$\pm$0.07     &
0.130$\pm$0.049         \\ [0.1cm]

$\eta_{1}$       &       0.0       &       0.0        & 0.24$\pm$0.35     &
 0.29$\pm$0.30            \\ [0.1cm]

$\delta$         &       0.0       &       0.0        & 17$\pm$28     &
 17$\pm$20            \\ [0.1cm]

$\xi_{0}$        & -0.53 $\pm$0.31  & -0.73$\pm$0.24  & 0.41$\pm$0.22     &
 0.42$\pm$0.16       \\ [0.1cm]

$\xi_{1}$        &       0.0        &      0.0      & 0.54$\pm$0.32 &
 0.53$\pm$0.17       \\ [0.1cm]

$\nu$       &        0.0       &      0.0      &  5.7$\pm$5.7  &
 4.3$\pm$2.7        \\ [0.1cm]

$\chi^{2}/N_{DOF}$ &   9.6/12      &      9.8/13     &     12.4/19       &
     13.0/20         \\ [0.1cm]  \hline

 \end{tabular}
\end{center}
\label{Tab_kp}
\end{table}
%
%
\section{The $\Lambda$ polarization in $\pi^{-}p$  and $\pi^{+}p$ collisions}
%
  The hyperon polarization data for $\pi^{\pm}p$ collisions have been analyzed
using eqs. (\ref{eq:Pol})- (\ref{eq:x2N}). The fit parameters are presented
in Table \ref{Tab_pi}. A cut $|x_{F}| \le 0.74$ is used to reduce exclusive
reaction contributions.
\subsection{The $\Lambda$ polarization in $\pi^{-}p$ collisions}
     The $\Lambda$ polarization in $\pi^{-}p$ collisions has been measured
at 3.95 GeV/c \cite{Adeva}, at 6 GeV/c \cite{Sugahara},
at 15 GeV/c \cite{Barreiro}, at 16.1 GeV/c \cite{Bensinger1} and
 at 18.5 GeV/c \cite{Stuntebeck}. The fit \# 1 is made with the $\omega$ 
parameter free and the fit \# 2 is performed with the $\omega=3.58$. 
The polarization is positive in the target fragmentation region and is 
negative or near zero for positive $x_{F}$.
\subsection{The $\Lambda$ polarization in $\pi^{+}p$ collisions}
     The $\Lambda$ polarization in $\pi^{+}p$ collisions has been measured
 at 18.5 GeV/c \cite{Stuntebeck}. The fit \# 3 is made with the $\omega$ 
parameter free and the fit \# 4 is made with the $\omega=3.58$.

  The polarization magnitude for $\Lambda$ hyperons produced in $\pi^{\pm}p$
collisions is smaller than it is in the $pp$ collisions.
\begin{table}[htb]
\small
\caption{Fit parameters of eqs.~(10)-(20) for $\Lambda$
 production in $\pi^{\pm}p$ collisions.}
\begin{center}
\begin{tabular}{cccccc}  \\[0.1cm]  \hline
Fit \#           &      1           &        2        &          3       & 
         4             \\[0.1cm] \hline  \\[0.1cm]

$Beam$              & $\pi^{-}$     & $\pi^{-}$        & $\pi^{+}$        &
 $\pi^{+}$          \\ [0.1cm]

$c$              & -1.8 $\pm$ 0.5   &-1.7$\pm$0.6    &-2.2$\pm$2.2      &
 -2.5$\pm$2.9        \\ [0.1cm]

$\omega$         & 4.33 $\pm$ 0.94  &   3.58         &5.2$\pm$6.5       &
   3.58              \\ [0.1cm]

$\sigma$        &  1.0             &   1.0        &    1.5      &
   1.5              \\ [0.1cm]

$p_{T}^{0}$     &        1.0       &        1.0       &     1.0           &
     1.0            \\ [0.1cm]

$\eta_{0}$      & 0.564 $\pm$ 0.088 & 0.632$\pm$0.081 & 0.45$\pm$0.46     &
0.64$\pm$0.10           \\ [0.1cm]

$\eta_{1}$       &  2.0$\pm$2.1  &  2.5 $\pm$1.0  &     0.0           &
     0.0            \\ [0.1cm]

$\delta$         &  50$\pm$35     &  49$\pm$16       &     0.0           &
     0.0            \\ [0.1cm]

$\xi_{0}$        &  0.276$\pm$0.044  & 0.264$\pm$0.048 &-0.14$\pm$0.70    &
-0.35$\pm$0.23       \\ [0.1cm]

$\xi_{1}$        & 0.45 $\pm$ 0.47  & 0.50 $\pm$0.43   &    0.9          &
   0.9           \\ [0.1cm]

$\nu$       &  12$\pm$11    &  13$\pm$11    &    1.0           &
    1.0         \\ [0.1cm]

$\chi^{2}/N_{DOF}$ &   20.9/19      &      21.2/20     &     0.40/3       &
      0.42/4         \\ [0.1cm]  \hline

 \end{tabular}
\end{center}
\label{Tab_pi}
\end{table}
%
%
\section{Polarization of  antihyperons produced using baryon and
 antibaryon beams}
%
   The reactions presented in this section can be considered as exotic ones
 due to a very unusual behaviour of the corresponding polarization with
 the rise of scaling variables. This behaviour cannot be predicted by the
existing models.   Most of the theoretical models predict zero polarization for
 antihyperon production in $pp(A)$ collision since they do not have  valence 
quarks common with the beam or target hadrons. The recent experiments have 
revealed non-zero polarization for  $\bar{\Xi}^{+}$ \cite{Ho} and  
 $\bar{\Sigma}^{-}$ \cite{Morelos} hyperons. Other antihyperons also
indicate non-zero polarizations with  small but not negligible magnitudes.

The fit parameters are presented in  Table \ref{T7_anti}. The $\alpha$
parameter in  eq. (\ref{eq:Pol}) is set to zero due to a limited statistics
and too few used targets.
\subsection{Polarization of $\bar{\Lambda}$ in $pA$ collisions}
 The $\bar{\Lambda}$ polarization have been measured in $pBe$ collisions
 at 400 GeV/c \cite{Heller_41,Lundberg} and at 800 GeV/c \cite{Ramberg}.

 Although the magnitude of the polarization is very
low ($\simeq 0.02$), the fit indicates an oscillation of the polarization
as a function of $x_{F}$ with good $\chi^{2}/N_{DOF}=0.81$.
 The data fits have been performed for $\omega$ parameter free (fit \# 1)
and $\omega=22.27$ (fit \# 2). Due to a limited statistics and kinematic range
of the data, the $\eta_{1}$ parameter is fixed at the same value (0.31)
as it is for $\Lambda$ polarization in $pA$ collisions. As we will see below
 the large value of $\omega=18.5 \pm 5.7$ is
typical for such exotic process ($p\rightarrow \bar{\Lambda}$)
 which corresponds to the $\Delta B=2$ exchange, where $B$ is a baryon
number. 
\subsection{Polarization of $\bar{\Xi}^{+}$ in $pA$ collisions}
 The $\bar{\Xi}^{+}$ polarization have been measured at 800
 GeV/c in collisions of a proton beam with a $Be$ target \cite{Ho}.
 The data fit has been performed $\omega=65.19$ (fit \# 3), which follows
from eq. (\ref{eq:omega}).
At the same time this value of $\omega$ is a minimal one which describes the
unusual polarization dependence on $x_{F}$ with the middle point closer to zero
 than the  other ones. The fit function indicates that polarization
 magnitude could be about 0.18. If such an unusual behaviour will be confirmed
by future experiments that reaction will be a large challenge to the strong
interaction theory. Much more data are desirable since we have only 
3 data points for this reaction.
\subsection{The $\bar{\Lambda}$ polarization  in $\bar{p}p$ collisions}
\begin{figure}[ht]
\centerline{\epsfig{file=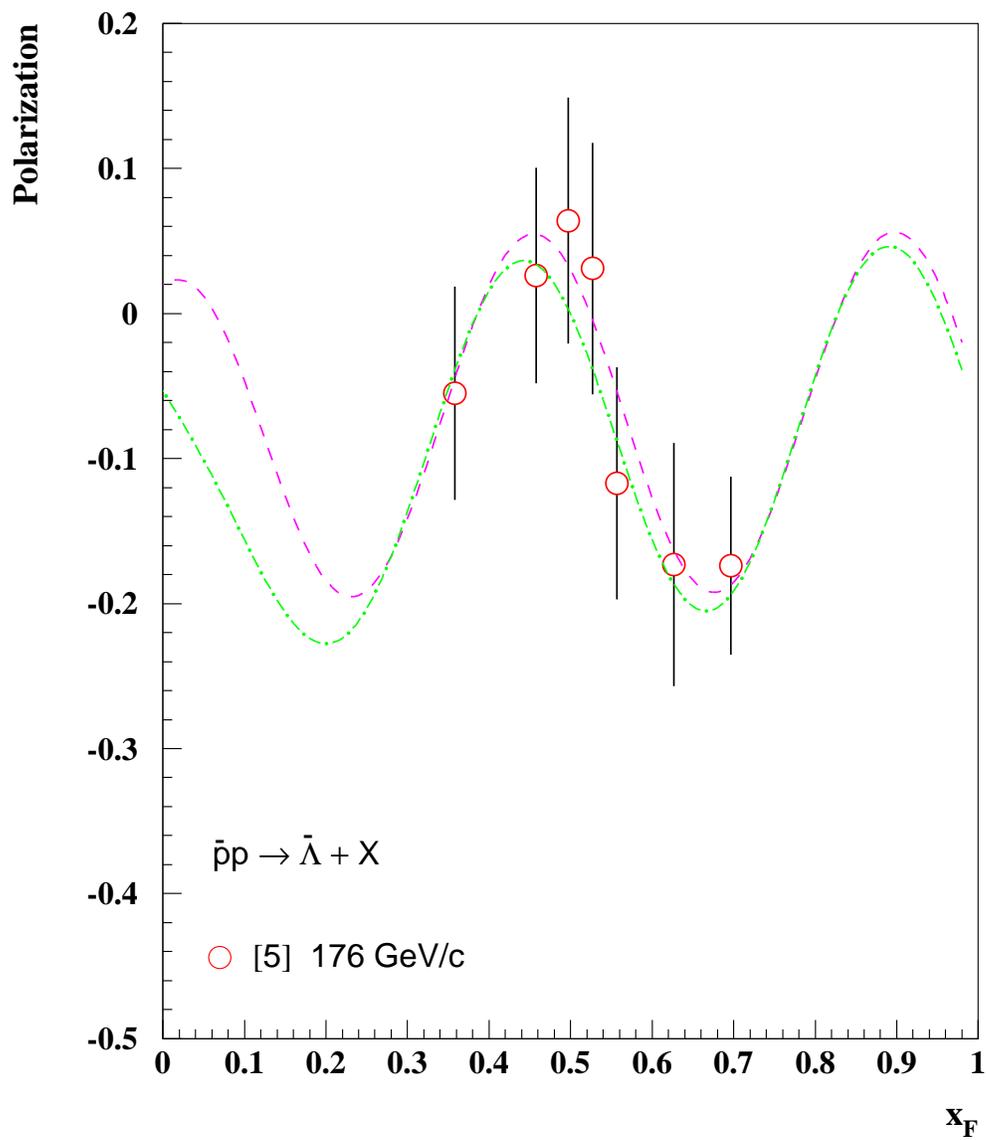,width=15cm}}
\caption {Polarization vs $x_{F}$ for $\bar{\Lambda}$ production 
in $\bar{p}p$ collisions. The fit parameters of eqs. (10)-(20)  are
presented in Table 7. The  curves correspond to the fit \# 5 for
$p_{T}=0.5$ GeV/c (dashed),  and $p_{T}=1.5$ GeV/c
(dash-dotted), respectively.}
    \label{lab_pm_xf}
\end{figure}

  The polarization of $\bar{\Lambda}$ ($\Pol_{\bar{\Lambda}}$) hyperons 
in $\bar{p}p$ collisions  has been studied at 176 GeV/c \cite{Gourlay}.
The fit \# 4 is made with $\omega$ parameter free, and the fit \# 5 is made for 
$\omega= 14.21$. The dependence of $\Pol_{\bar{\Lambda}}$
on $x_{F}$ is shown in  Fig. \ref{lab_pm_xf} along with fit \# 5 predictions
for $p_{T}=0.5$ GeV/c and $p_{T}=1.5$ GeV/c, respectively.  The predictions 
are made   for 176 GeV/c $\bar{p}p$ collisions.  Both fits indicate local  
maximums of $|\Pol_{\bar{\Lambda}}|$ 
at $x_{F} \approx 0.45$ and at $x_{F} \approx 0.65$ and an oscillation of
$\Pol_{\bar{\Lambda}}$ as a function of $x_{F}$ ($\omega = 16.2 \pm 4.1$). 
An interesting feature of
the data is that almost a full period of the oscillations is covered by the
data. This observation needs additional conformation in different kinematic
regions due to a limited statistics and the kinematic range covered by the
 data. Additional measurements are desirable to clarify
the $p_{T}$ dependence of the $\Pol_{\bar{\Lambda}}$.

The large values of $\omega$ have been found also for
 $p p(C) \rightarrow \bar{\Lambda} + X$
 and for $p A \rightarrow \bar{\Xi}^{+} + X$ reactions
 considered above. The dependence of the $\omega$ parameter on quantum
 numbers of the hadrons participating in the reaction will be discussed in
 details below.
\begin{table}[htb]
\small
\caption{Fit parameters of eqs.~(10)-(20) for reactions
  $pp \rightarrow \bar{\Lambda} + X$,
  $pp \rightarrow \bar{\Xi}^{+} + X$,
and  $\bar{p}p \rightarrow \bar{\Lambda} + X$.}
\begin{center}
\begin{tabular}{cccccccc}  \\[0.1cm]  \hline
Fit \#           &       1          &         2        &           3       & 
         4       &       5              \\[0.1cm] \hline \\[0.1cm]

$Process$  & $pA\rightarrow \bar{\Lambda}$ & $pA\rightarrow \bar{\Lambda}$  & 
$pA\rightarrow \bar{\Xi}^{+}$  &  
$\bar{p}A\rightarrow \bar{\Lambda}$  & $\bar{p}A\rightarrow \bar{\Lambda}   $ 
 \\ [0.1cm]

$c$              & -0.47 $\pm$0.23  & -0.49$\pm$0.22          &
 -16$\pm$12    & 4.5$\pm$2.9     & 3.5$\pm$1.2    \\ [0.1cm]

$\omega$         &  18.5 $\pm$5.7   &  $22.27$           &
 65.19            &  16.2$\pm$4.1    &  $14.21$            \\ [0.1cm]

$\sigma$        &      1           &       1          &
      1          &  -1             &   -1              \\ [0.1cm]

$p_{T}^{0}$     &        0.34       &      0.34        &
       0.4       &    0.6$\pm$0.9   &  0.4$\pm$0.6     \\ [0.1cm]

$\eta_{0}$       &  0.185$\pm$0.029 & 0.174$\pm$0.015   &
 0.045$\pm$0.007 &  0.071$\pm$0.061 & 0.042$\pm$0.016  \\ [0.1cm]

$\eta_{1}$       &       0.31       &      0.31         &
 0.0             &  0.0             &   0.0            \\ [0.1cm]

$\delta$         & 3.7$\pm$1.9      &   4.3$\pm$1.1      &
 0.0             &  0.0             &    0.0     \\ [0.1cm]

$\xi_{0}$        & 0.309 $\pm$0.072 & 0.322 $\pm$0.077     &
0.181$\pm$0.007   &  0.294$\pm$0.020 & 0.303$\pm$0.021  \\ [0.1cm]

$\xi_{1}$        & 0.07$\pm$0.15    & 0.10 $\pm$0.16       &
      0.0        &        0.0       &      0.0    \\ [0.1cm]

$\nu$       &       1.27       &       1.27          &
      0.0        &      0.0         &       0.0   \\ [0.1cm]

$\chi^{2}/N_{DOF}$ &    12.9/16     &      13.3/17      &
      0.00/0     &      0.74/2      &      1.08/3   \\ [0.1cm]  \hline

 \end{tabular}
\end{center}
\label{T7_anti}
\end{table}
\subsection{Polarization of $\bar{\Sigma}^{-}$ in $pCu$ collisions}
  The $\bar{\Sigma}^{-}$ polarization has been measured in the $pCu$
collisions at 800 GeV/c \cite{Morelos}. The  fit was not made since only 
four data points have been measured at fixed $x_{F}\approx 0.47$. The sign of 
polarization is positive and its magnitude is about $0.088 \pm 0.011$.
\section{Polarization of  $\Xi^{-}$ and $\Omega^{-}$ in 
 collisions of  hyperons  and protons with nuclei}
%
 In this section we analyze the polarization of $\Xi^{-}$ and $\Omega^{-}$
hyperons produced by a neutral unpolarized beam containing hyperons
and the $\Xi^{-}$ polarization produced by $\Sigma^{-}$ beam. 
In addition, the $\Omega^{-}$ and the proton polarizations in $pp(A)$ 
collisions are
 analyzed and compared with the hyperon polarization in different reactions.
The $\alpha$ parameter in  eq. (\ref{eq:Pol}) is set to zero due to 
a limited number of used targets and statistics.
\subsection{Polarization of $\Xi^{-}$ in  collisions
of $\Lambda$ and $\Xi^{0}$ with $Be$ target}
  In this subsection we consider the data on $\Xi^{-}$ 
polarization which have been measured using a neutral unpolarized
high energy beam containing $\Lambda$ and $\Xi^{0}$ hyperons \cite{Heller3}.
The primary 800 GeV/c proton beam was used to produce a neutral strangeness
containing beam, which in its turn interacted with a $Be$ target. The average
 momentum of produced $\Xi^{-}$ and $\Omega^{-}$ hyperons was about 395 GeV/c.
 This value of momentum was used to estimate the momentum of the neutral beam.
It is assumed in this analysis that the ratio ($R$) of the neutral beam
 momentum to the 800 GeV/c primary momentum is the same as the ratio of 
$\Xi^{-}$ momentum to the neutral beam momentum. These relations give 
$R=0.703$ and the neutral 
beam momentum about $R\cdot 800=562$ GeV/c. Using this assumption we 
performed a fit of the data \cite{Heller3}. The fit parameters are presented 
in Table \ref{Tab_str1}.

\begin{figure}[ht]
\centerline{\epsfig{file=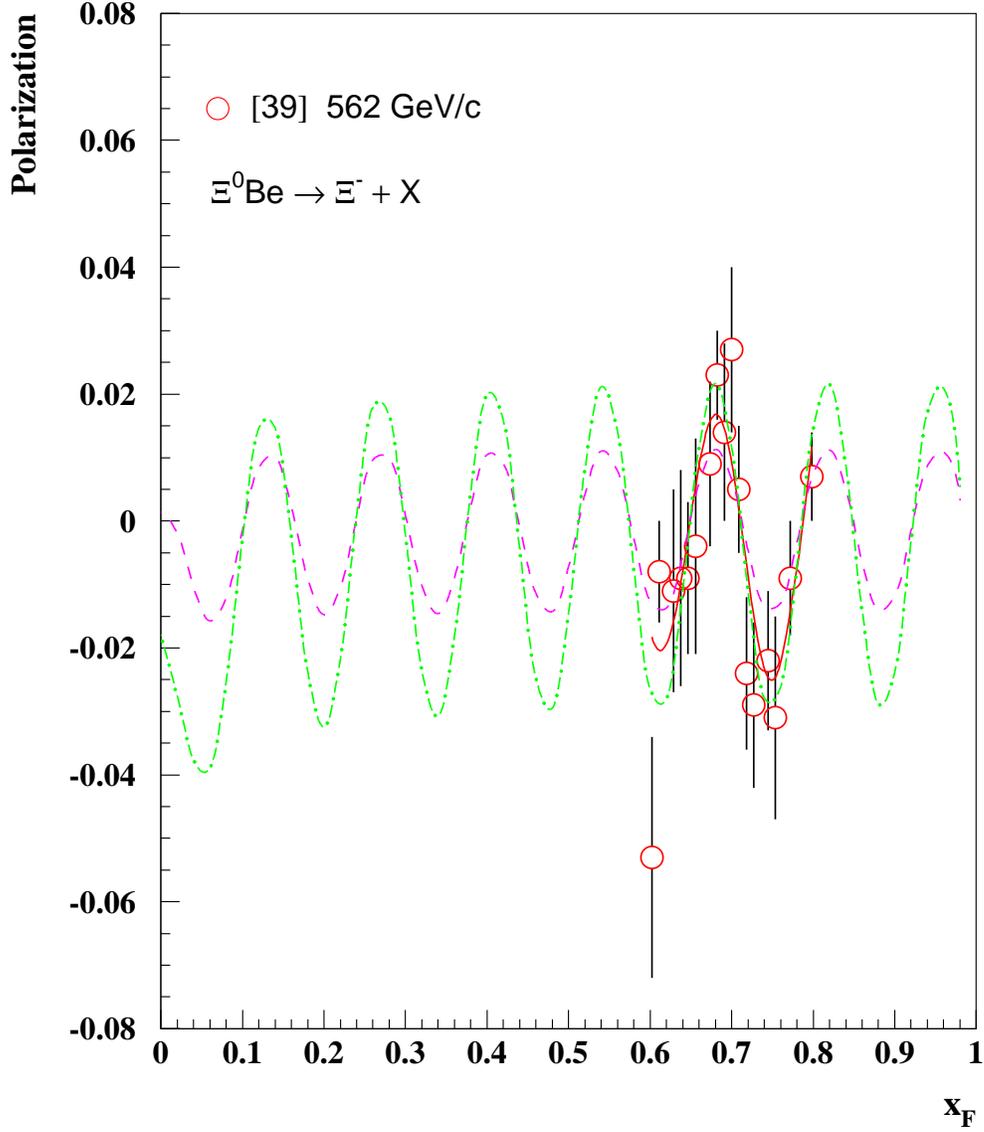,width=15cm}}
\caption {Polarization vs $x_{F}$ for $\Xi^{-}$ production in 
$\Lambda(\Xi^{0})p$ collisions. The fit parameters of eqs. (10)-(20) 
 are presented in Table 8. The  curves correspond to the fit \# 2 for
$p_{T}=0.5$ GeV/c (dashed),  and $p_{T}=1.5$ GeV/c
(dash-dotted), respectively.}
    \label{xim_xi0_xf}
\end{figure}

 The polarization of $\Xi^{-}$ was fitted with $\omega$ parameter free
 (fit \# 1) and with $\omega=45.97$ (fit \# 2). The $x_{F}$ dependence of
 polarization is shown in Fig. \ref{xim_xi0_xf} along with the fit \# 2 
predictions for $p_{T}=0.5$ GeV/c and $p_{T}=1.5$ GeV/c. Though the magnitude
 of polarization ($\Pol_{H}^{Max}$) is about 0.026 only, a clear $x_{F}$ 
dependence is seen, which is consistent with the oscillation behaviour, 
predicted by eqs. (\ref{eq:Pol})-(\ref{eq:x2N}). The frequency parameter 
$\omega = 46.0 \pm 3.7$ is very large and more than a full period of oscillation
is seen in Fig.  \ref{xim_xi0_xf}.
\subsection{Polarization of $\Omega^{-}$ in  collisions
of $\Lambda$ and $\Xi^{0}$ with $Be$ target}
 The $\Omega^{-}$ polarization has been measured using unpolarized neutral beam
 described in the previous subsection \cite{Heller3}.
 The polarization of $\Omega^{-}$ was fitted 
  with $\omega=6.090$ (fit \# 3). The magnitude
 of $\Omega^{-}$ polarization is about 0.11.
\subsection{Polarization of $\Omega^{-}$ in $pBe$   collisions}
\begin{figure}[ht]
\centerline{\epsfig{file=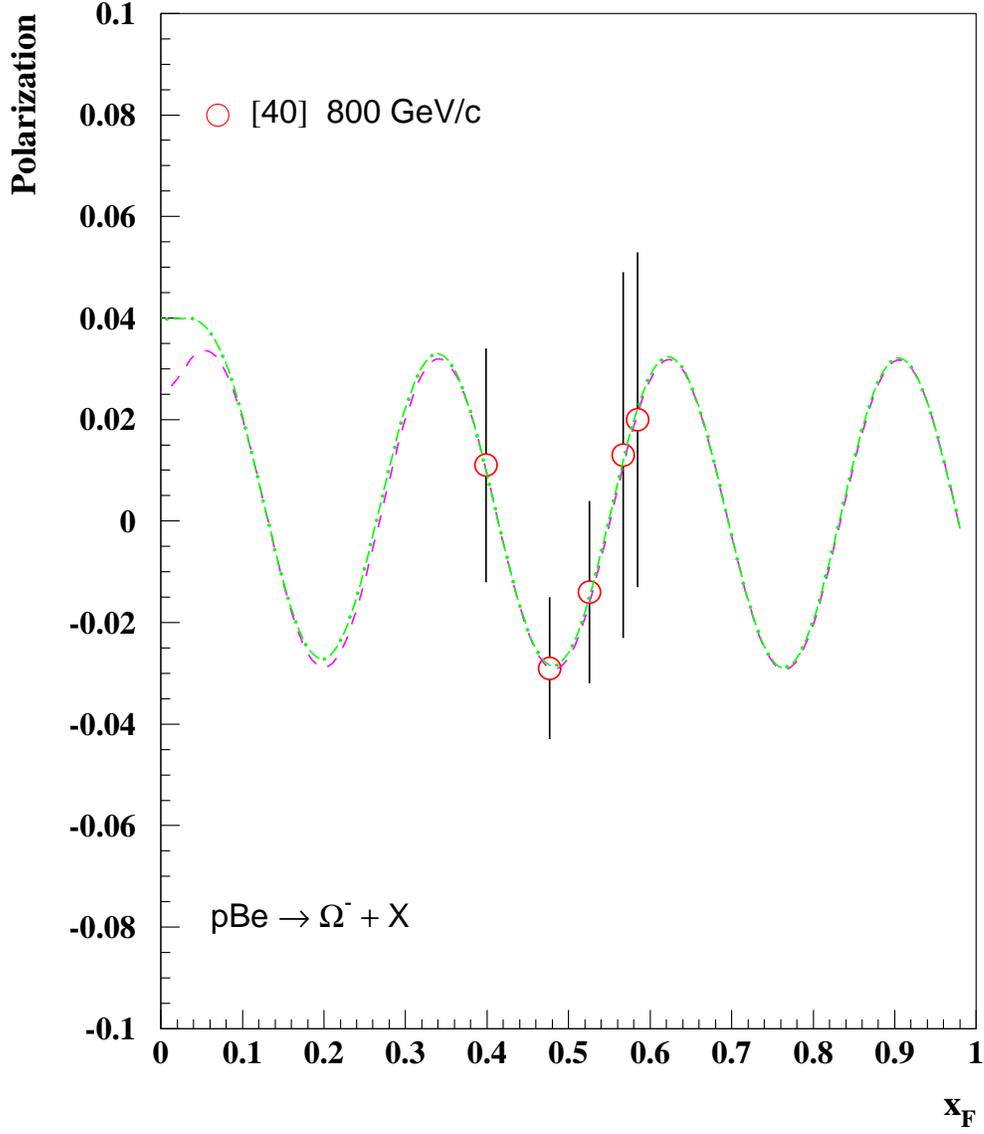,width=15cm}}
\caption {Polarization vs $x_{F}$ for $\Omega^{-}$ production in 
$pBe$ collisions. The fit parameters of eqs. (10)-(19)
 are presented in Table 8. The  curves correspond to the fit \# 4 for
$p_{T}=0.5$ GeV/c (dashed),  and $p_{T}=1.0$ GeV/c
(dash-dotted), respectively.}
    \label{omeg_pp_xf}
\end{figure}

 The $\Omega^{-}$ polarization has been measured using 800 GeV/c $pBe$
 collisions for $0.3<x_{F}<0.7$ and $0.5<p_{T}<1.3$ GeV/c \cite{Luk}.
 The mean value of polarization in this kinematic range is $-0.01 \pm 0.01$,
 but the dependence
 of it on $x_{F}$ clearly indicates its oscillation as a function
 of $x_{F}$ (see Fig. \ref{omeg_pp_xf}). The fit \# 4 is performed (see Table
\ref{Tab_str1}) with $\omega$ parameter fixed at 22.27 as is predicted 
by eq. (\ref{eq:omega}). It has to be mentioned that this large value of 
$\omega$ was estimated using the other reactions and at the same time 
it is quite consistent with the observed $x_{F}$ dependence for the 
$pBe \rightarrow \Omega + X$ reaction. The magnitude of polarization 
oscillation is about $0.032 \pm 0.016$.
\begin{table}[htb]
\small
\caption{Fit parameters of eqs.~(10)-(20) for
polarization of $\Xi^{-}$ and $\Omega^{-}$ hyperons in
collisions of a neutral unpolarized beam
containing $\Lambda$ and $\Xi^{0}$ with $Be$ target,
 and the $\Omega^{-}$ polarization in $pBe$ collisions.}
\begin{center}
\begin{tabular}{ccccccc}   \\[0.1cm] \hline
Fit \#           &      1          &        2        &          3       & 
         4             \\[0.1cm] \hline  \\[0.1cm]

$Process$       & $\Xi^{0} \rightarrow \Xi^{-}$     
                & $\Xi^{0} \rightarrow \Xi^{-}$        
                & $\Lambda \rightarrow \Omega^{-}$         
                & $p \rightarrow \Omega^{-}$         \\ [0.1cm]

$c$              & 2.37 $\pm$0.44  &2.37$\pm$0.42    & 
 2.6$\pm$4.1     & 1.36$\pm$0.69   \\ [0.1cm]

$\omega$         & 46.0  $\pm$3.7   & $45.97$        &
 $6.090$          & 22.27    \\ [0.1cm]

$\sigma$        &      -1.0         &     -1.0        &
  -1.0          & 1.0    \\ [0.1cm]

$p_{T}^{0}$     &        1.0        &       1.0       &
       1.0      & 0.39     \\ [0.1cm]

$\eta_{0}$       & 0.151$\pm$0.017  & 0.151$\pm$0.003 & 
 0.24$\pm$0.10 & 0.064$\pm$0.012     \\ [0.1cm]

$\eta_{1}$       &       0.0        &      0.0        & 
       0.0      & 0.0 \\ [0.1cm]

$\delta$         &       0.0        &      0.0        & 
       0.0       & 0.0      \\ [0.1cm]

$\xi_{0}$        & 0.085$\pm$0.023  &  0.085$\pm$0.002   & 
0.14$\pm$0.03    & 0.372$\pm$0.026 \\ [0.1cm]

$\xi_{1}$        &       0.0        &      0.0        & 
       0.0       & 0.0  \\ [0.1cm]

$\nu$       &       0.0        &        0.0      &    
       0.0       & 0.0    \\ [0.1cm]

$\chi^{2}/N_{DOF}$ &   13.75/13     &      13.75/14   &  
      0.04/1       & 0.01/2  \\ [0.1cm]  \hline

 \end{tabular}
\end{center}
\label{Tab_str1}
\end{table}
\subsection{Polarization of $\Xi^{-}$ in  collisions
of $\Sigma^{-}$  with $C(Cu)$ target}
 The polarization of $\Xi^{-}$ has been measured using 330 GeV/c $\Sigma^{-}$
 beam on  Carbon and  Copper targets \cite{Adamovich}.
 The polarization of $\Xi^{-}$ for a sample, combining the $C$ and the
 $Cu$ target measurements, was fitted with   free $\omega$ parameter
 (fit \# 1) and with $\omega=3.045$ (fit \# 2). The fit 
parameters  are presented in  Table 9.   The  $\Xi^{-}$
 polarization is negative and its magnitude is about 0.4.
\begin{table}[htb]
\small
\caption{Fit parameters of eqs.~(10)-(20) for
reactions $ \Sigma^{-} + C(Cu) \rightarrow \Xi^{-} + X$
and $p(C) + p \rightarrow p + X$.}
\begin{center}
\begin{tabular}{cccccc}   \\[0.1cm] \hline  
Fit \#           &       1        &          2       & 
         3      &      4                  \\[0.1cm] \hline  \\[0.01cm]

$Process$ & $\Sigma^{-}A \rightarrow \Xi^{-}$  
          & $\Sigma^{-}A \rightarrow \Xi^{-}$  
          & $pA \rightarrow p $
          & $pA \rightarrow p $         \\ [0.1cm]

$c$              &-1.53$\pm$0.56    &-1.51$\pm$0.48    
                 &  0.74 $\pm$0.14  & 0.73$\pm$0.10    \\ [0.1cm]

$\omega$         & 3.3$\pm$3.5      & $3.045$           
                 & 6.12 $\pm$0.21   & 6.090       \\ [0.1cm]

$\sigma$        & 0.87$\pm$0.27    & 0.88$\pm$0.23   
                 &      1.0         &       1.0       \\ [0.1cm]

$p_{T}^{0}$      &     1.0          &       1.0       
                 &       0.4        &       0.4    \\ [0.1cm]

$\eta_{0}$       & -0.41$\pm$0.51   &-0.446$\pm$0.042  
                 & 0.647 $\pm$0.014 & 0.648$\pm$0.012    \\ [0.1cm]

$\eta_{1}$       & -0.20$\pm$0.28   & -0.21$\pm$0.29   
                 & 0.204$\pm$0.045  & 0.205$\pm$0.045    \\ [0.1cm]

$\delta$         &     4.6          &      4.6       
                 &       1.75       &      1.75       \\ [0.1cm]

$\xi_{0}$        &  0.84$\pm$0.71   & 0.88$\pm$0.23   
                 & 1.474$\pm$0.089  & 1.476$\pm$0.090  \\ [0.1cm]

$\xi_{1}$        &  0.71$\pm$0.78    & 0.75$\pm$0.34   
                 &      1.75        &      1.75     \\ [0.1cm]

$\nu$       &      1.3         &      1.3     
                 &      1.75        &      1.75      \\ [0.1cm]

$\chi^{2}/N_{DOF}$ &   10.7/17      &      10.7/18      
                 &     35.9/31      &      35.9/32   \\ [0.1cm]  \hline

 \end{tabular}
\end{center}
\label{Tab_str2}
\end{table}
\subsection{The proton polarization in $pp(C)$ collisions}
  It is interesting to compare the hyperon polarization data with the proton
polarization data in $pp(C)$ collisions. The data have been measured at 100,
 200, 300 and 400 GeV/c in collisions of a proton beam with protons and 
a Carbon target \cite{Polvado}. The data fits have been performed for $\omega$ 
parameter free (fit \# 3) and $\omega=6.090$ (fit \# 4). 
The value of $\omega=6.12 \pm 0.21$ is consistent with the corresponding values
of  $\omega$ for $pA \rightarrow \Xi^{0}(\Xi^{-}) + X$ reactions, indicating
a similar mechanism for the polarization origin. The $p_T$ dependence of the 
polarization is mainly due to a significant $p_T$ dependence of the phases 
$x_{1}$ and $x_{2}$. The data are consistent with the scaling described by 
eq. (\ref{eq:Pol}) with ${\FPT}(p_{T})$ given by eq. (\ref{eq:F2}).
\section{Dependence of the hyperon polarization on quantum numbers}
  There is a very interesting feature of the hyperon polarization,
related with the value of the parameter $\omega$ in eqs. 
(\ref{eq:Pol})-(\ref{eq:x2}). The results of the data fits indicate
(see Table \ref{Pol_max}) that for
the hyperon and antihyperon production in $pp(A)$, $K^{\pm}p$, $\pi^{\pm}p$, 
$\Sigma^{-}p$,  $\Lambda(\Xi^{0})p$, and $\bar{p}p$ collisions
 the $\omega$ parameter depends on flavors of the projectile, the produced
hyperon and the target. It can be expressed for the reaction 
$a + b \rightarrow c^{\uparrow} + X$ using the  formula
\begin{equation}
\omega_{Q} = \sum_{i=1}^{i=5} a_{i}Q_{i},
\label{eq:omega}
\end{equation}
where $Q_{i}$ depends on the quark content of hadrons participating in 
the reaction, and $a_{i}$ are fit parameters. In particular, 
\begin{equation}
Q_{1} = |B_{c}|[n_{q}(a\bar{c}) + n_{q}^{ext}(a\bar{c})], 
\label{eq:Q1}
\end{equation}
\begin{equation}
Q_{2} = B_{a\bar{c}}\Psi(B_{a}) +
2\delta[n_{q}(a\bar{c})-6] \delta(B_{a\bar{c}} ), 
\label{eq:Q2}
\end{equation}
\begin{equation}
Q_{3} = |B_{a}||B_{c}| \{ B_{b\bar{c}} + 
2\delta[n_{q}(a\bar{c})-6] \delta(B_{b\bar{c}} ) \} , 
\label{eq:Q3}
\end{equation}
\begin{equation}
 Q_{4} = |B_{a}|\delta(N_{s}^{c}-2)
         \{\delta[n_{q}^{s}(a\bar{c})]\delta[n_{q}(a\bar{c})-2]
 + \delta[n_{q}^{s}(a\bar{c})-2]\},
\label{eq:Q4}
\end{equation}
\begin{equation}
 Q_{5} = \delta(B_{a}) \delta(B_{c}),
\label{eq:Q5}
\end{equation}
 where
\begin{equation}
 \Psi (n) =  \cases
 {n       ,  & if $n \ne 0$; \cr
  1       ,  & otherwise,\cr} \quad
\label{eq:Psi_n}
\end{equation}
and
\begin{equation}
 \delta (x) =  \cases {
 0,   & if $x \ne 0$; \cr
 1,   & otherwise. \cr}  \quad 
\label{eq:delta}
\end{equation}

\begin{figure}[ht]
\centerline{\epsfig{file=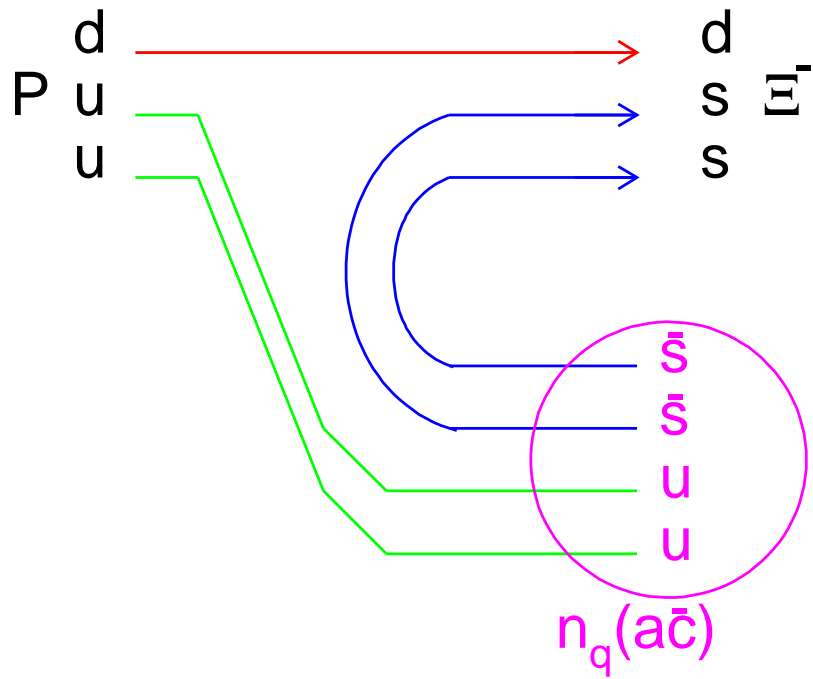,width=15cm}}
\caption {The diagram of a proton fragmentation into a $\Xi^{-}$ hyperon.
 The number of residual quarks is characterized by $n_{q}(a\bar{c})$. }
    \label{Diag1}
\end{figure}

In the above formulas $n_{q}(a\bar{c})$ is the minimal number of  quarks in the 
$a\bar{c}$ system ($q\bar{q}$ pairs of the same flavor are cancelled, see
quark diagram in Fig. \ref{Diag1}). The $n_{q}^{ext}(a\bar{c})$ denotes
additional (extra) quarks and antiquarks, produced in a process
above the minimal level when a higher
order quark level diagram is used, as in the case of the process 
$p + p \rightarrow p + X$. In the last mentioned process $n_{q}(a\bar{c})=0$,
 and we assume that the inclusive  protons are produced via a single valence
 quark fragmentation. So, four additional quarks and antiquarks are produced 
in this process, similar to the case of  $p + p \rightarrow \Xi^{-} + X$ 
process, shown in  Fig. \ref{Diag1}. For all other processes presented in
  Table \ref{Pol_max} the $n_{q}(a\bar{c}) > 0$ and $n_{q}^{ext}(a\bar{c})=0$.
 The sum $N_{spec} = [n_{q}(a\bar{c}) + n_{q}^{ext}(a\bar{c})]$ can also 
be considered as the number of spectator quarks in a process $a \rightarrow c$.

We may conclude from the analysis of 22 reactions presented in Table
10 that non-zero number of spectator quark $N_{spec}$ is always
needed (which means quark exchange diagram) to generate a non-zero baryon
polarization. 

The $n_{q}^{s}(a\bar{c})$ is the net 
number of $s$ quarks or antiquarks in the $a\bar{c}$
system. The  $B_{a\bar{c}}$  and $B_{b\bar{c}}$ are the baryon numbers of
 the $a\bar{c}$  and the $b\bar{c}$ systems, respectively. The $B_{a}$,
$B_{b}$ and  $B_{c}$ are the corresponding baryon numbers, and $N^{c}_{s}$
is the number of $s$ quarks in a produced hadron $c$.
 
 The value $n_{q}(a\bar{c})$ is used in the Constituent Interchange model (CIM)
\cite{CIM1}
which predicts a  cross section at $p_{T}=0$ of the form
\begin{equation}
 { {Ed^{3}\sigma} \over{d^{3}p} }(a\rightarrow c) \sim 
 (1-x_{F})^{2n_{q}(a\bar{c})-3}.
\label{eq:CIM}
\end{equation}
The first term of eq. (\ref{eq:omega}) contributes to the processes
$B + B \rightarrow B + X$, where $B$ means baryon
( reactions \# 1 - \# 5), the second one takes into account the specific
properties of the meson induced reactions \# 6 - \# 12, the third term is 
important for the antibaryon production, the $Q_{4}$ takes into account very 
high ``oscillation frequency'' data for the (anti)hyperons, containing  $two$ 
$s$ quarks when the beam hadron has $two$ ($zero$) $s$ quarks (see reactions
 \# 15-16).  

The term $Q_{5}$ is introduced to take into account the reactions
$meson + p(A) \rightarrow meson + X$, for which all other terms are zero.
The preliminary analysis of analyzing power for these reactions gives 
$a_{5} \approx 12$ (see also \cite{Prep98}). A detailed study of the meson
induced reactions will be made in a separate paper.

The reactions in the Table \ref{Pol_max} can be classified according to
the set of $Q_{i}$ values.
The most simple is the case of baryon production in $pp(A)$ collisions when
only the $Q_{1}$ term is non-zero and the $\omega \approx 1.5 N_{spec}$ is
related with the number of spectator quarks as is illustrated by a quark
diagram in Fig. \ref{Diag1}.  We may suggest that the mean colour field
created by the spectator quarks in these reactions is the reason of 
polarization oscillation as a function of $x_{F}$.

The $\omega$ values presented  in  Table \ref{Pol_max} were fitted using 
eq. (\ref{eq:omega}) and the parameters $a_{1}-a_{4}$ are shown
in  Table \ref{omega_fit}. The reactions \#11,15,18 and 20-22
were not used in the fit, since they have  too few data points.

 If the inclusive protons are produced via the single beam quark fragmentation
 similar to the reaction \# 2, then for the number of
 spectator quarks $n_{q}^{ext}(\bar{a}c)=4$  the predicted value 
$\omega_{Q}=6.090$ is in a good agreement with the estimated value 
$6.12\pm0.21$. 

\begin{figure}[ht]
\centerline{\epsfig{file=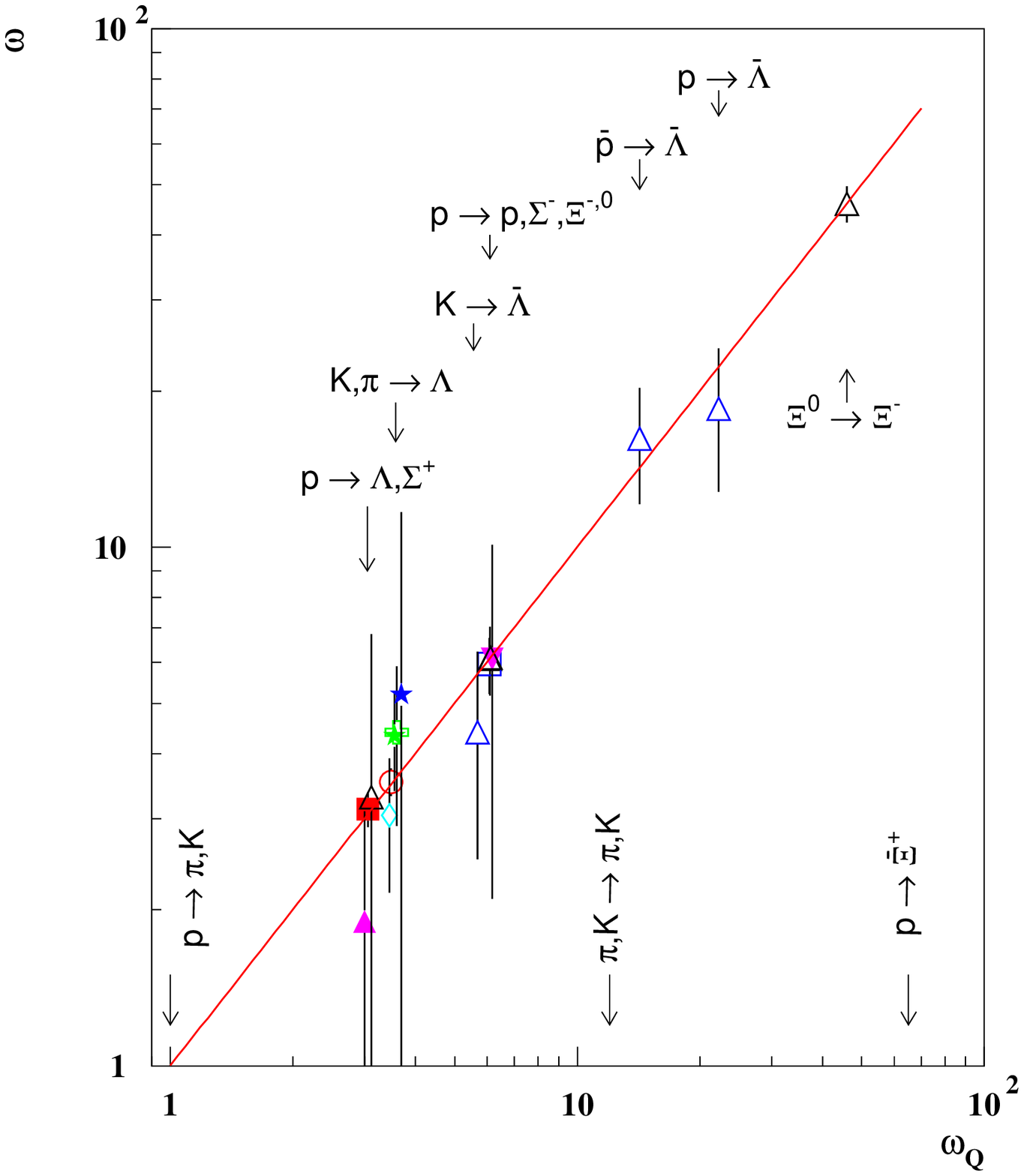,width=15cm}}
\caption {The estimated $\omega$ parameter vs the predicted one, $\omega_{Q}$.}
    \label{ome_vs_omeq}
\end{figure}

 The dependence of the $\omega$ parameter vs $\omega_{Q}$ is shown in 
 Fig. \ref{ome_vs_omeq} for 15 different reactions, presented in 
Table \ref{Pol_max}. The arrows indicate also the predictions
 of the $\omega=0.98$  for the analyzing power in the reactions 
$p^{\uparrow}p \rightarrow \pi (K) + X$, $\omega \approx 12$  in the reactions
$meson + p^{\uparrow} \rightarrow meson + X$, and $\omega \approx 65.2$ for
 polarization in the reaction $pp \rightarrow \bar{\Xi}^{+} + X$.

\begin{table}[Htb]
\small
\caption{The comparison of the estimated $\omega$ parameter and its
 prediction $\omega_{Q}$ from eq. (22).
The estimate of maximum in hyperon polarization magnitude
 using the fit parameters of  eq.~(10)-(20). The maximum is estimated
 for the $x_{F} \ge 0 $ region. }
\begin{center}
\begin{tabular}{cccccccccc}    \\[0.1cm] \hline  \\[0.1cm]
 \#  &       reaction  & $Q_1$ & $Q_2$ & $Q_3$ & $Q_4$  &   $\omega$   &
 $\omega_{Q}$ & ${\Pol_{H}^{max}}$ &   ${P_{Q}}$
    \\[0.1cm] \hline \\[0.1cm]
 1& $p Be  \rightarrow \Lambda$   & 2 & 0 & 0 & 0 & 3.13$\pm$0.24 &
 3.045 & -0.272$\pm$0.011     &  -0.275   \\ [0.1cm]

 2& $p Be  \rightarrow \Xi^{0}$   & 4 & 0 & 0 & 0 & 6.10$\pm$0.92 &
 6.090  & -0.141$\pm$0.052     &  -0.138   \\ [0.1cm]

 3& $p Be  \rightarrow \Xi^{-}$   & 4 & 0 & 0 & 0 & 5.95$\pm$0.74 &
 6.090  & -0.140$\pm$0.021     &  -0.138   \\ [0.1cm]

 4& $p Be  \rightarrow \Sigma^{+}$ & 2 & 0 & 0 & 0 & 1.9$\pm$1.2   &
 3.045  &  0.337$\pm$0.076     &   0.275   \\ [0.1cm]

 5& $p Be  \rightarrow \Sigma^{-}$ & 4 & 0 & 0 & 0 & $6.1\pm$4.0   &
 6.090  &  0.34$\pm$0.26       &   0.275  \\ [0.1cm]

 6& $K^{-}p \rightarrow \Lambda$  & 3 &-1 & 0 & 0 & $3.53\pm$0.22 &
 3.58  &  0.61$\pm$0.11       &   0.604  \\ [0.1cm]

 7& $K^{-}p \rightarrow \Xi^{-}$  & 3 &-1 & 0 & 0 & $3.04\pm$0.88 &
 3.58  &  0.53$\pm$0.17       &    0.604  \\ [0.1cm]

 8& $K^{+}p \rightarrow \Lambda$  & 3 &-1 & 0 & 0 & $4.4\pm$1.5   &
 3.58  & $\pm 0.23\pm$0.15    &   -0.052  \\ [0.1cm]

 9& $\pi^{-}p \rightarrow \Lambda$ & 3 &-1 & 0 & 0 & $4.33\pm$0.94 &
 3.58  & $\pm 0.23\pm$0.14    &   -0.178  \\ [0.1cm]

10& $\pi^{+}p \rightarrow \Lambda$ & 3 &-1 & 0 & 0 & $5.2\pm$6.5   &
 3.58  & $\pm 0.14\pm$0.17    &   -0.178  \\ [0.1cm]

11 & $K^{-}p \rightarrow \bar{\Lambda}$  & 3 & 1 & 0 & 0 &       &
 5.55  &    0.4$\pm$0.4     &    0.302  \\ [0.1cm]

12& $K^{+}p \rightarrow \bar{\Lambda}$  & 3 & 1 & 0 & 0 & $4.4\pm$1.9 &
 5.55  &   0.78$\pm$0.25      &    0.604   \\ [0.1cm]

13& $pBe \rightarrow \bar{\Lambda}$ & 6 & 2 & 2 & 0 & $18.5\pm$5.7 &
22.27  & $\pm 0.021\pm$0.010  &   -0.025  \\ [0.1cm]

14& $\bar{p}Be\rightarrow \bar{\Lambda}$ & 2 & 0 & 2 & 0 & 16.2$\pm$4.1 &
14.21  & $\pm 0.187\pm$ 0.046 &    0.194  \\ [0.1cm] 

15& $pBe \rightarrow \bar{\Xi}^{+}$ & 6 & 2 & 2 & 1 &              &
 65.19 & $\pm 0.18\pm$0.03       &   -0.188   \\ [0.1cm]

16& $\Xi^{0} Be \rightarrow \Xi^{-}$ & 2 & 0 & 0 & 1 &  46.0$\pm$3.7 &
45.97  & $\pm 0.026\pm$0.005  &    0.022  \\ [0.1cm]

17& $\Sigma^{-} C \rightarrow \Xi^{-}$ & 2 & 0 & 0 & 0 & 3.3$\pm$3.5 &
 3.045  & -0.4$\pm$0.4       &   -0.275   \\ [0.1cm]

18& $\Lambda Be\rightarrow \Omega^{-}$ & 4 & 0 & 0 & 0 &          &
 6.090 &  0.111$\pm$0.067     &    0.132   \\ [0.1cm]

19& $pp(C) \rightarrow  p        $ & 4 & 0 & 0 & 0 & 6.12$\pm$0.21  &
 6.090 & $\pm 0.072\pm$0.014  &    0.072   \\ [0.1cm]

20& $p Be \rightarrow \Omega^{-}$ & 6 & 2 & 2 & 0 &       &
 22.27  &  0.032$\pm$0.016    &     0.031   \\ [0.1cm]

21& $p Cu \rightarrow \bar{\Sigma}^{-}$ & 6 & 2 & 2 & 0 &    &
 22.27  &  0.088$\pm$0.011    &     0.092   \\ [0.1cm]

22& $p Be  \rightarrow \Sigma^{0}$ & 2 & 0 & 0 & 0 &  &
 3.045 &  0.28$\pm$0.13       &     0.275   \\ [0.1cm]
\hline
 \end{tabular}
\end{center}
\label{Pol_max}
\end{table}
 The line in  Fig. \ref{ome_vs_omeq} shows the result of the fit 
$\omega = r\cdot \omega_Q$ with $r=1.002 \pm 0.025$
 and $\chi^{2}/N_{DOF}=0.38/16$.
 This figure confirms a strong correlation of the $\omega$ parameter 
with the quantum numbers which characterize the reaction.

The magnitude of hyperon polarization (${\Pol_{H}^{max}}$) varies 
significantly
 with the hyperon, the projectile and the target flavors and the target atomic
weight. The values of ${\Pol_{H}^{max}}$ for different reactions were estimated
for  $x_{F} \ge 0$, where most of the data have been measured,
 using eqs. (\ref{eq:Pol})-(\ref{eq:x2N}) and fit parameters, taken from 
 Tables 1-9. The sign of the ${\Pol_{H}^{max}}$
 is indicated the same as that  for the experimental data if it does not 
vary in the mentioned above region, and $\pm$ - otherwise.   The results 
are presented in  Table 10 for 22 different reactions. 

It is easy to notice from  Table \ref{Pol_max} that there is 
a correlation between the value of ${\Pol_{H}^{max}}$ and the $\omega$ 
parameter  for the corresponding reactions. The product of these two values, 
$|{\Pol_{H}^{max}}|\cdot \omega$ varies much less than
each value separately.
\begin{table}[Htb]
\small
\caption{The  fit parameters of  eqs. (21) and (32). }
\begin{center}
\begin{tabular}{ccccccccc}    \\[0.1cm] \hline 
   $a_{1}$    &    $a_{2}$    &    $a_{3}$    &    $a_{4}$    
   &  $a_{5}$ & $\chi^{2}/N_{DOF}$ \\[0.1cm] \hline 
1.523$\pm$0.045 & 0.98$\pm$0.24 & 5.6$\pm$1.70 & 43.0$\pm$3.7 
   & $\approx 12$ & 3.57/12  \\ [0.1cm] \hline
   $b_{1}$    &    $b_{2}$    &    $b_{3}$    &    $b_{4}$    
   & $b_{5}$  & $\chi^{2}/N_{DOF}$ \\[0.1cm] \hline
0.70$\pm$0.19 &0.545$\pm$0.066 & 4.94$\pm$0.73 & 11.6$\pm$1.5 
   & -4.73$\pm$0.74   & 5.5/16 \\ [0.1cm] \hline
 \end{tabular}
\end{center}
\label{omega_fit}
\end{table}

  The $\sigma$ parameter  was found to be consistent with 
\begin{equation}
 \sigma_{Q} =  \cases
 {  1       ,  & for $pp(A)$ collisions; \cr
   -\Phi_Q  ,  & otherwise,\cr}
 \quad
\label{eq:lambda_Q}
\end{equation}
where
\begin{equation}
 \Phi_Q =  (-1)^{n_{c}} 
 \Psi(B_{c})\Psi(Y_{a}) \Psi(Y_{c})/\Psi(\Delta S_{ac}),
\label{eq:Phi_Q}
\end{equation}
where  $Y_{a}$, $Y_{c}$, and $\Delta S_{ac}=S_{a}-S_{c}$   are hypercharges 
($Y=B+S$) of $a$, $c$, and strangeness change, respectively. 
The $n_{c}$ is the number of quarks with parallel spins  in the
 hadron $c$. It is assumed here, in accordance with 
the $SU(6)$ quark model \cite{Lach,Franklin}, that
 $n_{c}=3$ for the $\Omega^{-}$, $n_{c}=2$ for the $\Xi^{0,-}$, 
$\Sigma^{\pm,0}$,  protons, neutrons, and $n_{c}=1$ for the $\Lambda$.
 
  The magnitude ${\Pol_{H}^{max}}$ of the polarization can be approximated
 by
\begin{equation}
 P_{Q} = {\epsilon \Phi_{Q}
                   \over{1 + \Delta \omega(\hat{b})  } },
\label{eq:P_Q}
\end{equation}
where the value of  $\Delta \omega(\hat{b})$ is given by eq.
\begin{equation}
\Delta \omega (\hat{b}) = \sum_{i=1}^{i=5} b_{i}Q_{i}^{'},
\label{eq:D_omega}
\end{equation}
where
\begin{equation}
Q_{1}' = |B_{c}| n_{q}^{ext}(a\bar{c}), 
\label{eq:q1}
\end{equation}
\begin{equation}
Q_{2}' = \delta(B_{a}) [\Delta S_{ac} - S_{a}S_{c}]
+2\delta(S_{c}+3)\delta(S_{a}+1)
\label{eq:q2}
\end{equation}
\begin{equation}
Q_{3}' = |B_{a}||B_{c}| \{ B_{b\bar{c}}/\Psi[4(n_{c}-N_{s}^{c})+1] + 
\delta[n_{q}(a\bar{c})-6] \delta(B_{b\bar{c}} ) \} , 
\label{eq:q3}
\end{equation}
\begin{equation}
 Q_{4}' = B_{c}|B_{a}|\delta(N_{s}^{c}-2)
         \{\delta[n_{q}^{s}(a\bar{c})]\delta[n_{q}(a\bar{c})-2]
 + \delta[n_{q}^{s}(a\bar{c})-2]\},
\label{eq:q4}
\end{equation}
\begin{equation}
 Q_{5}' = B_{b\bar{c}} \delta(B_{a\bar{c}} ).
\label{eq:q5}
\end{equation}
The $\epsilon=0.275 \pm 0.010$  and $b_{i}$ are the fit 
parameters, shown in Table 11.

 The fit of $|{\Pol_{H}^{max}|}$  using $|P_{Q}|$ (the sign of 
${\Pol_{H}^{max}}$ was not taken into
 account) was made for the reactions \# 1 - \# 22.
 The values of $P_{Q}$ are presented in Table \ref{Pol_max}  with the sign 
in the region $x_{F}>0$ given by eq.  (\ref{eq:P_Q}).

The magnitude of polarization in the reaction $pp \rightarrow \Lambda + X$
is $-\epsilon$, see eq. (\ref{eq:P_Q}). For the $\Xi^{-}$ and $\Xi^{0}$ hyperon
production in $pp$ collisions the magnitude of polarization is $-\epsilon/2$ 
due to the factor $1/\Psi(\Delta S_{ac})$ in eq. (\ref{eq:Phi_Q}).

  The $p_{T}^{0}$ parameter  was found to be consistent with an approximation
\begin{equation}
 p_{T}^{0} = k_{T}^{0}/[n_{q}(a\bar{c})+ n_{q}^{ext}(a\bar{c})],
\label{eq:pt0}
\end{equation}
 where $k_{T}^{0}= 2.32 \pm 0.14$ GeV/c.

Eqs. (\ref{eq:omega}) and (\ref{eq:P_Q}) should have 
their explanations in the models of strong interaction and 
ultimately in the QCD, perhaps, in the non-pertubative approach.

The value of parameter $\omega$ can also be estimated for the processes 
in which the analyzing power is measured. That requires additional precise 
measurements in a wide range of $x_{F}$ in a kinematic area where the 
$x_{A}$ scaling is fulfilled ($p_{T}\ge$ 1 GeV/c and/or the beam energy 
$\ge 40$ GeV).  The reactions with meson  beams are of special 
interest because of a high value of the $\omega$ parameter and a possibility 
of ``oscillation''  of the analyzing power as  a function of $x_{F}$  
(compare Figs. \ref{lab_pm_xf}- \ref{omeg_pp_xf}
 and the preliminary results in \cite{Prep98}).

   If we assume that eqs. (\ref{eq:Pol})-(\ref{eq:omega}) are valid also
for the analyzing power ${\AN}$, we can make predictions of the $\omega$ for
 different reactions.
 
 In particular, we expect $\omega=0.98$ for the
$\pi^{\pm,0}$, $K^{\pm,0}$, $\eta$, $K^{*}(890)$, $\rho$, $\omega$
 production in  $pp$ and $\bar{p}p$
collisions; $\omega=3.045$ for the $\Lambda$ production in $pp$ collisions;
 $\omega=6.090$ for the proton production in  $pp$ collisions,
 $\omega \approx 12$ for $meson + p^{\uparrow} \rightarrow meson + X$, and
 $\omega=22.27$ for the antiproton and $\bar{\Lambda}$ production in $pp$ 
collisions. These predictions include about two tens different reactions and
a preliminary analysis confirms these predictions. We may conclude that
eq. (\ref{eq:omega}) has a high predictive power which allows one to predict
the $\omega$ parameter for about four tens different spin-dependent processes.
 It is highly unlikely that the $\omega$ value in so many different
processes is described by the same eq. (\ref{eq:omega}) 
  for accidental reasons.
Moreover, within the experimental errors we may fix two out of five parameters:
$a_{1}=3/2$, and $a_{2}=1$.

 Eq. (\ref{eq:omega}) also predicts that in some cases a target particle
plays a significant role in the dynamics of a hyperon production. This is
the case when the baryon number $B_{b\bar{c}}$ and $Q_{3}$ are different 
from zero.

 The scaling behaviour of the ${\AN}$ and ${\Pol_{H}}$
indicates that the corresponding processes take place at the quark
or parton level.
In case of the discussed above scaling for the analyzing power and for 
the hyperon polarization such constituents could be the constituent quarks
or the current quarks (see related discussion in \cite{Boros1}).
In order to resolve quark degrees of freedom inside a hadron, the transverse
momentum $p_{T}$ in a process should be higher than $p_{T}^{0}$. That could be
 the reason for the $p_{T}$ dependence of polarization (\ref{eq:F2}).

%

\section{Discussion}
 The results of the previous sections indicate that the use of two scaling
variables $x_{A\pm} =(x_{R}\pm x_{F})/2$ is essential for the universal energy
independent description of the existing hyperon polarization data in inclusive
reactions. The scaling variables transform into each other under rotation
 transformation around the normal to the scattering plane
and allow one to satisfy in a natural way the feature
(\ref{eq:Pasym}), which is also related with the rotation invariance.
The variables $x_{A\pm}$ treat on a more equal basis the data for the central
 and the fragmentation regions as well as the transverse  and the longitudinal
 momentum components of the produced hyperon.

The form of eq. (\ref{eq:Pol}) is chosen from the dimensional analysis, the
 rotation invariance requirement  and it is also motivated by the existing
data for both the hyperon polarization and the analyzing power in inclusive
 reactions for hadron-hadron collisions. The derived formulas are applicable
 for both the ${\AN}$ and the ${\Pol_{H}}$ data approximation and reflect 
the scaling properties of these two classes of processes at high enough 
energies and $p_{T}$. The specific range of energies and $p_{T}$ at which 
the scaling properties are valid could depend on the process type. 
The existence 
of the above scaling implies that large spin effects will survive at high 
energies and do not depend directly on the beam energy. These properties 
of the ${\AN}$ and the ${\Pol_{H}}$ are waiting for their explanation in the 
strong interaction theory in general and in the Quantum Chromodynamics 
(QCD) in particular.
\subsection{ Analogy between the scaling properties of hyperon polarization
and the analyzing power}
  It is important  to take into account in the future models a deep
 analogy between the hyperon polarization and  the analyzing power, which
follows from the results of this study and \cite{Prep98,EPJC,Liang1}.
 Below we list  some common features of the ${\AN}$ and the ${\Pol_{H}}$,
which  indicate  a common nature of both phenomena.

 Features of the analyzing power in inclusive hadron production by polarized
protons and antiprotons and features of the hyperon polarization in inclusive
hadron-nucleon collisions:

\begin{itemize}
\item[\bf {a)}] Scaling behaviour for the hyperon polarization
and the analyzing power as a function of
$x_{A\pm}=(x_{R}\pm x_{F})/2$  and $p_T$ in the beam fragmentation region,
and in  the central region.
\item[\bf {b)}] The ${\Pol_{H}}$ and the $A_{N}$ are approximated by a product
 of functions of $p_{T}$  and  $x_{A\pm}$:
${\Pol_{H}}={\FPT}(p_{T})[{\GXA}(x_{A+}-x_{2})-\sigma{\GXA}(x_{A-} +x_{2})]$,
 where the $x_{2}$  could depend on $p_{T}$  and a
 reaction type.
\item[\bf {c)}] The  function ${\FPT}(p_{T})$ 
 rises with $p_{T}$ at small $p_{T} \le p_{T}^{0} $ and have a plateau or
 decrease above $p_{T}^{0}$, where $p_{T}^{0}=$ 0.4-2 GeV/c depends on 
a reaction type.
\item[\bf {d)}] The  function ${\GXA}(x_{A\pm})$   is proportional to
$\sin [\omega(x_{A\pm} - x_{1}) ]$, where   $x_{1}$ could depend on $p_{T}$.
The $\omega$ is a universal function of quantum numbers characterizing the
reaction, common for the ${\Pol_{H}}$ and $A_{N}$.
\item[\bf {e)}] The ${\Pol_{H}}$ and the $A_{N}$ are zero at $p_T$ = 0 due to 
the absence of a preferable direction, such as a normal 
to the scattering plane. This implies ${\FPT}(0)=0$.
\item[\bf {f)}]The sign and the magnitude of the ${\Pol_{H}}$ and the $A_{N}$
 depend on the projectile, the target, and the produced hadron flavors. 
\end{itemize}

  We may conclude that a close similarity of the kinematic behaviour of
${\Pol_{H}}$ and  $A_{N}$ and a universal dependence of the $\omega$
parameter on quantum numbers indicate on a common origin of spin effects in
these two classes of spin-dependent reactions.
\subsection{Interference origin of the hyperon polarization}
 The polarization of hadrons is a pure Quantum mechanics effect related
with the interference of spin-flip ($g$) and spin-nonflip ($f$) amplitudes.
The transverse hadron polarization can be expressed via the $f$ and $g$ 
amplitudes as \cite{Ryskin1}
\begin{equation}
   {\Pol_{H}} = 2Im(f^{*}g)/(|f|^{2} + |g|^{2}).
\label{eq:Im_fg}
\end{equation}
Taking into account eqs. (\ref{eq:Pol})-
(\ref{eq:GX}) we can choose amplitudes $f$ and $g$ as
\begin{equation}
  f \propto f_{0}\bigl\{ 
                      exp[-i\omega (x_{A+} - x_{1} - x_{2})/2] +
  (\sqrt{\sigma})^{*}exp[-i\omega (x_{A-} - x_{1} + x_{2})/2] 
\bigr\}, 
\label{eq:f}
\end{equation}
\begin{equation}
  g  \propto g_{0}{c\over{2\omega}} A^{\alpha} {\FPT}(\PT{})
  \bigl\{ 
                     exp[i\omega(x_{A+} - x_{1} - x_{2})/2] - 
 \sqrt{\sigma}exp[i\omega(x_{A-} - x_{1} + x_{2})/2] 
\bigr\},
\label{eq:g}
\end{equation}
where $f_{0}$ and $g_{0}$ are the functions of kinematic variables ($x_{F}$,
$p_{T}$, $\sqrt{s}$)  with zero relative phases,
 which have to satisfy the constrains followed from both the polarization
  and the cross section data. The generalized optical 
theorem predicts the following relation for a cross section:
\begin{equation}
  {\Pol_{H}}d\sigma = 2Im(f^{*}g),
\label{eq:cross}
\end{equation}
where $d\sigma$ is the corresponding unpolarized inclusive cross section
\cite{Soffer}, which can be used together with eq. (\ref{eq:Im_fg}) 
to fix the $f_{0}$ and $g_{0}$.

To have a non-zero  value of the ${\Pol_{H}}$ both amplitudes have to be 
non-zero and the phase difference $\Delta \phi$ between spin-flip and 
spin-nonflip amplitudes has to be non-zero too. For the cases of 
$\sigma = \pm 1$ the following phase differences are expected from 
eqs. (\ref{eq:f})-(\ref{eq:g}):
\begin{equation}
  \Delta \phi = (1 +\sigma) \pi/4 + \omega(x_{R}w_{1} -x_{1}) + Arg(c/\omega).
\label{eq:Dfg}
\end{equation}
 As we can see from eq. (\ref{eq:Dfg}), the variables $x_{R}$ and $\omega$ 
play an important
role in the hyperon polarization phenomena since they determinate the phase
difference between the spin-flip and non-flip amplitudes. 
The higher is the $\omega$ value, the larger  the $\Delta \phi$ change 
rate with the $x_{R}$ increase. Eq. (\ref{eq:omega}) for $\omega_{Q}$ can
 be considered as a sum of effective ``charges'', which create a mean field 
and lead to the change of the phase difference $\Delta \phi$.

In the lowest-order pertubative $QCD$ all amplitudes are relatively
real. This tends to rule out polarization in the hard scattering of partons,
which seems to be well described in the low-order $QCD$. The observation of
undiminished polarization near $p_{T}=4$ GeV/c implies that either pertubative
 $QCD$ does not apply or that another mechanism is responsible, such as 
interference of exited states or the fragmentation process \cite{Lundberg}.
\subsection{Some theoretical ideas for hyperon polarization}
 Several phenomenological models have been proposed to explain the hyperon
 polarization data and the analyzing power data (see recent review in
 \cite{Soffer}). Some of the models  have the features that allow one to 
understand, at least at a qualitative level, the analogy between 
the ${\AN}$ and the ${\Pol_{H}}$ discussed above.

\subsubsection{Orbital motion of valence quarks}
One class of such models assumes that an orbital motion of  valence quarks
 and  surface effects are responsible for the correlation between the quark
 spin direction and transverse motion of produced hadron \cite{Liang1,Boros1}.
 By taking the $\Lambda$'s containing two, one or zero valence quark(s) of
 a beam proton into account, the model predicts the sign and the $x_{F}$
 dependence of  $\Lambda$'s polarization. The model predicts correctly
the sign and the magnitude of $\Sigma^{-}$, $\Xi^{0,-}$ polarization in
$pp$ collisions, and $\Lambda$ polarization in $K^{-}p$ collisions.
It also predicts a smaller magnitude for $\Lambda$'s produced by 
$\pi^{\pm}$ beams. The  authors of the model  claim an analogy of mechanisms,
 which lead to non-zero hyperon polarization and analyzing power.
The model does not explain the specific features of antihyperon polarization.
 
\subsubsection{Parton rotation inside constituent quarks}
   A separate approach was developed by Troshin and Tyurin, which assumes
the rotation of a quark-antiquark cloud inside constituent quarks
 \cite{Tyurin1,Tyurin2}. The main role belongs to the orbital
angular momentum and polarization of the strange quark-antiquark pairs in 
the internal structure of constituent quarks. The hyperons are produced
in two stages. At the first stage the overlapping and interaction of peripheral
clouds occur which results in massive quark appearance and a mean field is
generated. Constituent quarks located in the central part of hadron are 
supposed to scatter in a quasi-independent way by this mean field.
At the second stage  two mechanisms  take place:  Recombination 
of the constituent quarks with a virtual massive strange quark 
(soft interaction)
into a hyperon or a scattering of a constituent quark in the mean field,
excitation of this constituent quark, appearance of a strange quark
as a result of decay of the constituent quark and a subsequent fragmentation
of a strange quark into a hyperon (high $p_{T}$'s hard interaction).
The resulting expression at $p_{T}> 1$ GeV/c when hard interactions dominate is
\begin{equation}
 P(s,x,p_{T}) \simeq \sin[{\sl P_{q}} <L_{\bar{q}q}> ],
\label{eq:Trosh}
\end{equation}
where ${\sl P_{q}}$ is the polarization of the constituent quark $q$ which
arises due to multiple scattering in the mean field and $<L_{\bar{q}q}>$ is 
the mean value of an internal angular momentum inside the constituent quark.

Thus, in this model the polarization of a strange quark is the result of 
multiple scattering of a parent constituent quark, the correlation between 
the polarization 
of a strange quark and the polarization of the constituent quark and a local
compensation of a spin and an orbital angular momentum of a strange quark.

 The simplest possible $x$ dependence of ${\sl P_{q}}$ is taken
\begin{equation}
    {\sl P_{q}}(x) = {\sl P_{q}}^{max} x,
\label{eq:Trosh1}
\end{equation}
where ${\sl P_{q}}^{max} = -1$.

The model predicts the negative sign and $x_{F}$ dependence of the
$\Lambda$ polarization. Eq. (\ref{eq:Trosh}) resembles
 eq. (\ref{eq:GX}), especially in the beam fragmentation region,
with effective $\omega = <L_{\bar{q}q}>$.
Eqs. (\ref{eq:Trosh}), (\ref{eq:Trosh1}) predicts a scaling behaviour
of the hyperon polarization.
The concept of the mean field, generated by quarks, which leads to a hyperon
polarization and to its change with $x$ rise similar to eq. \ref{eq:GX}
 is also in  consent with the analysis, presented above.

There are no predictions for other hyperons, though the authors assume zero
polarization in inclusive process $pp \rightarrow p + X$ due to a low
probability of multiple scattering in the mean field in comparison with
a single scattering. A single scattering does not polarize quarks and protons
appear unpolarized in the final state since a single scattering is dominant in
this process.

  There are several semiclassical models, which provide simple arguments for
a qualitative description of the hyperon polarization, but since they fully
ignore the relevance of the phase difference, which is crucial, they are 
unable to make solid quantitative predictions.
\subsubsection{The recombination model}
 In  the recombination model  \cite{DeGrand1,DeGrand2}
  a dynamical reason for the  spin-momentum correlation is
explained  by the effect of Thomas precession
\cite{Thomas,Logunov}. The effect arises when the direction of the force
acting on a quark does not coincide with the direction of its motion. It
leads to a rotation of the quark spin and could be the reason of the
discussed above ``oscillation'' of polarization or analyzing power as a 
function of $x_{F}$. The Thomas frequency is an inverse function of a quark
mass
\begin{equation}
  \overrightarrow{\omega}_{T} = 
{ {\gamma} \over{\gamma +1}} { \overrightarrow{F}\over{m_{q}} } 
\times \overrightarrow{V},
\label{eq:thom1} 
\end{equation}
where $V$ is the strange quark velocity, $F$ - the force, $m_{q}$ - 
the strange quark mass, and $\gamma = (1-V^{2})^{-1/2}$. An additional term
 will appear in the effective Hamiltonian which describes the recombination 
process
\begin{equation}
  U = \overrightarrow{S} \cdot \overrightarrow{\omega}_{T},
\label{eq:thom2}
\end{equation}
where $\overrightarrow{S}$ is a spin of the quark. Within the old-fashioned 
 perturbation theory the final expression for the $\Lambda$ polarization is
\begin{equation}
  P(p\rightarrow \Lambda) = - {  {12p_{T}x_{F} (x_{F} - 3 x_{s}) } 
  \over{  \Delta x_{0} M^{2} (x_{F} + 3x_{s})^{2} } }   ,
\label{eq:thom3}
\end{equation}
where it is assumed that a recombination time
$\Delta t \approx (p^{ave}_{z}/m_{q}) \Delta x_{0}$, the average
momentum of the quark is $p^{ave}_{z} \approx P(x_{F}+3x_{s})/6$ and
$\Delta x_{0} \approx 4$ GeV$^{-1}$ is a distance scale of the order of 
the proton radius. The $M\simeq 2$ GeV/c$^2$ is an effective mass and
the $x_{s}$ is a fraction of a proton momentum ($P$) which carries
the $s$ quark \cite{DeGrand1}.
These assumptions lead to a quark mass cancellation in the polarization
formula (\ref{eq:thom3}) and a scaling behaviour of the $\Lambda$ polarization.
The model gives the right sign and a good approximation of the $x_{F}$ 
dependence for the $\Lambda$ polarization. There are also many predictions for 
hyperon polarization in other reactions. They are based on some rules which 
are formulated within the framework of the recombination model. In particular,
 there is a statement that the effect of recombination of the partons in 
the proton as they are transferred into the outgoing hadron may be 
different depending on
whether they are accelerated (as are the slow sea partons) or decelerated
(as are the fast valence partons). It is also a statement that two partons with
similar wave functions in the proton may interact with themselves differently
not as they interact with a parton whose wave function is different. This
results in a simple rule: slow partons preferentially recombine with their 
spins down in the scattering plane while fast partons recombine with their
spins up.

  The model predicts correctly the polarization sign but not the magnitude
for some of the reaction. So, the model predicts the same polarization 
magnitude for $p \rightarrow \Lambda$ as it is  for $p \rightarrow \Xi^{-,0}$,
but we know from the above analysis that the polarization magnitudes for 
$\Xi^{-,0}$ are two times smaller. Similarly, for $K^{-}p \rightarrow \Lambda$
process the model predicts the same magnitude as for $p \rightarrow \Lambda$,
while the measured value is two times larger. Since the polarization in the
model is essentially kinematic at the quark level all the antibaryons should
 have zero polarization. As we have seen in the above analysis, the
experimental situation is much more complicated.

 There is another estimate of the hadronization time which follows from the
analysis of $A$-dependence of hadron production
$\Delta t \approx p^{c}/M_{0}^{2}$, where 
$M_{0} \approx 1$ GeV \cite{Hadron_t}. The use of this $\Delta t $ estimate 
results in a different expression for the $\Lambda$ polarization
\begin{equation}
  P(p\rightarrow \Lambda) = - {  {2p_{T} M_{0}^{2} x_{F} (x_{F} - 3 x_{s}) } 
  \over{  m_{q} M^{2} x_{R} (x_{F} + 3x_{s}) } }   ,
\label{eq:thom4}
\end{equation}
which is an inverse function of the $s$ quark mass $m_{q}$. This example
shows that the hyperon polarization is sensitive to the details of the
hadronization process. These results  seem to imply that the origin
of hyperon polarization is closely related with the confinement mechanism.

  Since the Thomas precession frequency is an inverse function of the quark
mass $m_{q}$, this may be the reason for a large variation of the $\omega_{Q}$
parameter for different reactions. The ratio $m_{s}/m_{d}$ is estimated to be
from 17 to 25 with a mean about 21 \cite{PPB}. The same
order of magnitude is given by the ratio $65.19/3.045=21.4$ of the 
$\omega_{Q}$ parameters for the processes $p \rightarrow \bar{\Xi}^{+}$ and
$p \rightarrow \Lambda$, in which $s$ and $\bar{d}$ quarks play an important 
role. It is interesting to estimate the total rotation angle of a quark 
spin using the above approximations
\begin{equation}
 \phi_{rot} = \omega_{T} \Delta t \approx
{  { p_{T} \cdot  (x_{F} - 3 x_{s})} \over
 { m_{q} \cdot  (x_{F} + 3x_{s})} }.
\label{eq:N_turn}
\end{equation}
Taking typical $x_{F}=0.7$, $x_{s}=0.1$, $p_{T}=1$ GeV/c, and 
$m_{s}= 122$ MeV/c$^{2}$, we have $\phi_{rot}=3.3$, while for the $d$ quark
with $m_{d}= 6$ MeV/c$^{2}$ the $\phi_{rot}=67$. So, the total rotation angle
 of a quark spin due to the Thomas precession could be rather large and
varies in the same range as that of the  $\omega_{Q}$. 

In case of $\Xi^{-}$ and $\Xi^{0}$ production in $pp$ collisions the effective
field created by the spectator quarks $\propto n_{q}(a\bar{c}) $ is expected to
be two times higher than for $\Lambda$ production and the corresponding 
rotation angle will be also two times larger, in agreement with the $\omega$ 
value for $\Xi^{-}$ and $\Xi^{0}$ production.

\subsubsection{Lund model}
 Another explanation of spin-momentum correlation follows from a picture 
of a colour flux tube, which emerges after the collision between an outgoing
 quark and the rest of hadronic system \cite{Ander,RYSK}.
The $SU(6)$ wave function is assumed for hadrons, in particular, for $\Lambda$
the $(ud)$ system is in a singlet state, so the $\Lambda$ polarization is
that of the $s$ quark. An outgoing $ud$ diquark with spin $S=0$ and isospin
$I=0$ stretches the colour field and a $s\bar{s}$ pair is produced. It is
assumed that the $s$ quark has $p_{T}$ which must be locally compensated
by that of the $\bar{s}$ quark. As a result, the $s\bar{s}$ pair has an 
orbital momentum which is assumed to be balanced by the spin of the $s\bar{s}$
pair. The model predicts a negative $\Lambda$ polarization in $pp$ collisions
but cannot predict its magnitude or $x_{F}$ dependence. The $p_{T}$ dependence
of the polarization is linear. The model needs additional assumptions
to explain the polarization in other reactions and fails to explain
the antihyperon polarization.

\subsubsection{Optical approximation}
 We  propose a simple toy model which uses an analogy with the optics. Let us
 consider the $\Xi^{-}$ production in a collision of two protons ($a$ and $b$)
in their c.m. reference frame. The proton's longitudinal size is about
$2R_{h}/\gamma_{cm}$, where $R_{h} \approx 0.8$ fm is a proton radius and
 $\gamma_{cm}=E_{cm}/(2m_{p}c^{2})$.

In an optical picture the phase can be related with the number of scattering
centers \cite{UFN}.
  We assume here that a hadron can be characterized by an effective
refractive index ($n$) which leads to a phase difference 
$\chi = (n-1)d \cdot p_{q}/\hbar$ between
spin-flip and spin-nonflip  quark scattering amplitudes, where $d$ is a total
path length inside a proton and $p_{q}\approx p_{c}/z \approx p_{a}x_{R}/z$ is
a quark momentum. 
 It is assumed here for the sake of simplicity that a quark from the proton $a$
passes on average half of the proton $b$ thickness and then changes its angle
due to a scattering in the proton $b$. The second part of its way inside 
the proton is approximately by a factor of $1/\cos{\theta_{cm}}=x_{R}/x_{F}$ 
larger than that  before the scattering (we consider here not too large 
scattering angles). This results in a phase difference
\begin{equation}
 \chi_{a} \approx  
{  R_{h} (n-1) x_{R} \over {\lambda_{p} <z> x_{F} } }
(x_{F} + x_{R} ),
\label{eq:phase}
\end{equation}
where $\lambda_{p}=\hbar/(m_{p}c) \approx 0.210 $ fm is the proton
 Compton  wavelength. Eq. (\ref{eq:phase}) can be rewritten as 
$\chi_{a} = \omega_{eff}\cdot x_{A+}$, where 
\begin{equation}
 \omega_{eff} =  
{ 2 R_{h} (n-1) x_{R} \over {\lambda_{p} <z> x_{F} } },
\label{eq:omeg_eff}
\end{equation}
and $<z>$ is the mean fraction of a quark momentum which is carried by
the produced hyperon.
A similar consideration of the proton $b$ quark scattering inside the
proton $a$ results in a $\chi_{b} = \omega_{eff}\cdot x_{A-}$ and the total
contribution into the $\Xi^{-}$ polarization is
\begin{equation}
 {\Pol}_{\Xi-} \propto
 [\sin(\omega_{eff} \cdot x_{A+}) - \sin(\omega_{eff} \cdot x_{A-})],
\label{eq:pol_eff}
\end{equation}
which is very similar to  eq. (\ref{eq:Pol}). The averaging over the
 transverse quark coordinates inside a proton is not taken into account for 
the sake of simplicity. A more careful consideration of a quark path length
after the scattering removes the singularity $1/x_{F}$ in 
eq. (\ref{eq:pol_eff}) since the path length is limited at 
$\theta_{cm}=\pi/2$ by the $R_{h}$.
 The $p_{T}$ dependence of a hyperon polarization 
is also not taken into account in  eq. (\ref{eq:pol_eff}) since we assume 
that $p_{T}$ is high enough to resolve quarks inside the hadron structure. 
The condition for that is $p_{T} \ge  \hbar/r_{q} \approx
 3\hbar/R_{h} \approx 0.75$ GeV/c,  where $r_{q}$ is a constituent quark 
radius.  Comparison of eqs. (\ref{eq:omeg_eff}) and 
(\ref{eq:omega}) assumes that 
$(n-1) \propto \omega_{Q} \propto n_{q}(a\bar{c})$.

 We may learn from this toy model
that the hyperon polarization oscillation is probably related with 
a corresponding scattering amplitude phase change due to the hadron mean field
 generated during hadron interaction. 
The field strength is proportional to the number of scattering centers
$n_{q}(a\bar{c})$ in agreement with the above  analysis of experimental data.
The scaling variables $x_{A+}$ and 
$x_{A-}$ arise in this model from the consideration of geometrical and 
relativistic  properties of interacting hadrons  and the assumption that
the phase difference $\chi$ is proportional to the quark path length in 
the mean hadron field.

\section{Conclusion}
The analysis of experimental data on the hyperon and antihyperon polarization
 has been made.
  It is shown that the existing (anti)hyperon polarization data in inclusive
reactions for $pp(A)$, $\pi^{\pm}p$, $K^{\pm}p$, $\bar{p}p$ and 
hyperon-nucleon collisions can be described by a function of 
$p_{T}$ and two scaling variables $x_{A\pm} = (x_{R} \pm x_{F})/2$:
 $ {\Pol_{H}} = A^{\alpha} {\FPT}(p_{T})[{\GXA}(x_{A+} -x_{2})
        -\sigma {\GXA}(x_{A-} +x_{2}) ]$. 
 The function  ${\GXA}(x_{A\pm})$ is approximated 
in the scaling limit by a simple expression 
$\propto \sin[\omega(x_{A\pm} - x_{1})]$ which results in an oscillation of 
polarization in case of large $\omega$ value.
Functions ${\FPT}(p_{T})$, $x_{1}$,
and $x_{2}$ depend on $p_{T}$ and
can be approximated by constants above 1-2 GeV/c. 

It is found that the $\omega$ parameter is consistent with an universal
dependence on quark composition (eq. (\ref{eq:omega})) for about two tens
measured (anti)baryon production processes. In addition, the preliminary
analysis indicates the validity of the same eq. (\ref{eq:omega}) for
about two dozen spin-dependent processes in which analyzing power of produced
hadrons was measured. Universal character of the $\omega$ dependence on quantum
numbers allows one to predict the $\omega$ values for a large number of 
processes in which analyzing power or hadron polarization is measured. It 
indicates also that we deal with a physical case and not just with data 
description. 

In particular, we expect a near linear dependence of the analyzing power on 
$x_{F}$ for $\pi, K$ production in $p^{\uparrow}p$ and $\bar{p}^{\uparrow}p$ 
collisions due to a small value of $\omega \approx 0.98$. 

 The polarization of $\bar{\Lambda}$ in $\bar{p}p$  collisions is fitted well
by  eq. (\ref{eq:Pol}) and indicates an oscillation of it 
 as a function of $x_{F}$. Similar oscillations with large 
$\omega$ parameter  are also seen for other processes, including 
$\Xi^{0} \rightarrow \Xi^{-}$ and $p \rightarrow \Omega^{-}$. 

 The data fits indicate a simple relation for the $\omega$ parameter  
in case of hyperons produced in $pp$ 
collisions: $ \omega \approx   1.5 n_{q}(a\bar{c})  $,
 where  $n_{q}(a\bar{c})$ is the number of quarks in the $a\bar{c}$ system.

The data fits indicate also that the $\Lambda$ hyperon polarization decreases
 on nuclear targets according to the law 
${\Pol_{H}} \propto A^{\alpha_{1}|x_{F}|}$, with $\alpha_{1} \approx -0.16$.
 This effect may be related with the rescattering of polarized $s$ quarks in 
the nuclear matter before formation of a hyperon is over. Since the formation
 length  is proportional to the final hyperon momentum, we expect a rise
 of $s$ quark  rescattering probability with $|x_{F}|$ increase. Due to the
similarity between the hyperon polarization and the analyzing power 
 features we expect that the $A$-dependence of the analyzing power is 
also described by the law ${\AN} \propto A^{\alpha_{1}|x_{F}|}$ 
with $\alpha_{1} \approx -(0.1 \div 0.3)$.

The polarization sign and its magnitude  depend 
on quark composition of hadrons participating in the reaction and can 
be predicted using the proposed formulas. In most cases the processes with large
$\omega$ value have a small magnitude which makes it difficult to measure.

 There is an analogy between the scaling properties of polarization for 
hyperons, produced in collisions of unpolarized hadrons, and the scaling 
properties of the analyzing power of hadrons, produced in the collisions 
of polarized protons (antiprotons) with hadrons. This analogy between 
the analyzing power and  the hyperon polarization indicates on a common 
origin of both phenomena.
 The possible mechanism may be related with the confinment forces. 
These forces are created by the mean colour field generated after
the initial hard scattering of quarks. The quark spin rotation in the
mean field could lead to the phase difference between  spin-flip
and spin-nonflip amplitudes which results in the hyperon polarization.

The origin of the scaling variables $x_{A\pm} = (x_{F} \pm x_{R})/2$ can be
understood from the dimensional analysis, the rotational invariance, and also 
in the framework of an optical picture of hadron-hadron interactions.
%

\newpage
%
\newpage
\clearpage
\clearpage
\clearpage
\clearpage
\clearpage
\clearpage

\clearpage
\clearpage

\clearpage
\clearpage
\clearpage
\clearpage

\newpage
\begin{thebibliography}{99}
    \bibitem{Bunce} {\it}
G. Bunce et al., {\em Phys. Rev. Lett.\/} {\bf 36} (1976) 1113.
    \bibitem {Abe_D34}{\it}
F. Abe et al., {\em Phys. Rev.\/} {\bf D34} (1986) 1950.
    \bibitem{Abe_PRL50} {\it}
F. Abe et al.,  {\em Phys. Rev. Lett.\/} {\bf 50} (1983) 1102.
    \bibitem{Bonner} {\it}
B.E. Bonner et al., {\em Phys. Rev.\/} {\bf D38} (1988) 729.
    \bibitem{Dukes} {\it}
E.C. Dukes et al., {\em Phys.Lett.\/} {\bf B193} (1987) 135.
    \bibitem{Raych} {\it}
K. Raychaudhuri et al., {\em Phys. Lett.\/} {\bf 90 B} (1980) 319.
    \bibitem{Gourlay} {\it}
S.A. Gourlay et al., {\em Phys. Rev. Lett.\/} {\bf 56} (1986) 2244.
    \bibitem{Wilk_46} {\it}
C. Wilkinson et al., {\em Phys. Rev. Lett.\/} {\bf 46} (1981) 803.
    \bibitem{Heller_51} {\it}
K. Heller et al., {\em Phys. Rev. Lett.\/} {\bf 51} (1983) 2025.
    \bibitem{Heller_41} {\it}
K. Heller et al., {\em Phys. Rev. Lett.\/} {\bf 41} (1978) 607.
    \bibitem{Lundberg} {\it}
B. Lundberg et al., {\em Phys. Rev.\/} {\bf D 40} (1989) 3557.
    \bibitem{Ramberg} {\it}
E.J. Ramberg et al., {\em Phys. Lett.\/} {\bf 338 B} (1994) 403.
    \bibitem{Smith} {\it}
A.M. Smith et al., {\em Phys. Lett.\/}  {\bf 185} (1987) 209.
%
    \bibitem{Morelos} {\it}
 A. Morelos et al., {\em Phys. Rev. Lett.\/} {\bf 71} (1993) 2172.
%
    \bibitem{Wilk} {\it}
 C. Wilkinson et al., {\em Phys. Rev. Lett.\/} {\bf 58} (1987) 855.
%
    \bibitem{Ankenb} {\it}
 C. Ankenbrandt et al., {\em Phys. Rev. Lett.\/} {\bf 51} (1983) 863.
    \bibitem{Wah} {\it}
 Y.W. Wah et al., {\em Phys. Rev. Lett.\/} {\bf 55} (1985) 2551.
%
    \bibitem{Deck} {\it}
 L. Deck et al., {\em Phys. Rev.\/} {\bf D 28} (1983) 1.
%
    \bibitem{Bonner1} {\it}
B.E. Bonner et al., {\em Phys. Rev. Lett.\/} {\bf 62} (1989) 1591.

    \bibitem{Duryea} {\it}
 J. Duryea et al., {\em Phys. Rev. Lett.\/} {\bf 67} (1991) 1193.
%
    \bibitem{Ho} {\it}
P.M. Ho et al., {\em Phys. Rev. Lett.\/} {\bf 65} (1990) 1713.
%
    \bibitem{Trost} {\it}
 L.H. Trost et al., {\em Phys. Rev.\/} {\bf D 40} (1989) 1703.
%
    \bibitem{Rameika} {\it}
 R. Rameika et al., {\em Phys. Rev.\/} {\bf D 33} (1986) 3172.
%

    \bibitem{Faccini} {\it}
M.L.Faccini-Turleur et al., {\em Z. Phys.\/} {\bf C 1} (1979) 19.
%
    \bibitem{Borg} {\it}
 A.Borg et al., {\em Nuovo Cimento\/}  {\bf 22 A} (1974) 559.
%
    \bibitem{Abramowicz} {\it}
 H. Abramowicz et al., {\em Nucl. Phys.\/}  {\bf B 105} (1976) 222.
%
    \bibitem{Grassler} {\it}
 H. Grassler et al., {\em Nucl. Phys.\/}  {\bf B 136} (1978) 386.
%
    \bibitem{Baubillier} {\it}
 M. Baubillier et al., {\em Nucl. Phys.\/}  {\bf B 148} (1979) 18.
%
    \bibitem{Bensinger} {\it}
 J. Bensinger et al., {\em Nucl. Phys.\/}  {\bf B 252} (1985) 561.
%
    \bibitem{Ganguli} {\it}
 S.N. Ganguli et al., {\em Nucl. Phys.\/}  {\bf B 128} (1977) 408.
%
    \bibitem{Chliapnikov} {\it}
 P.V. Chliapnikov et al., {\em Nucl. Phys.\/} {\bf B 112} (1976) 1.
%
    \bibitem{Barletta} {\it}
 W. Barletta et al., {\em Nucl. Phys.\/} {\bf B 51} (1973) 499.
%
    \bibitem{Ajinenko} {\it}
I.V. Ajinenko et al., {\em Phys. Lett.\/} {\bf 121 B} (1983) 183.
%
    \bibitem{Adeva} {\it}
 B. Adeva et al., {\em Z. Phys.\/} {\bf C 26} (1984) 359.
%
    \bibitem{Sugahara} {\it}
 R. Sugahara et al., {\em Nucl. Phys.\/} {\bf B 156} (1979) 237.
%
    \bibitem{Barreiro} {\it}
 F. Barreiro et al., {\em Phys. Rev.\/} {\bf D 17} (1978) 669.
%
    \bibitem{Bensinger1} {\it}
 J. Bensinger et al., {\em Phys. Rev. Lett.\/} {\bf 50} (1983) 313.
%
    \bibitem{Stuntebeck} {\it}
P.H. Stuntebeck et al., {\em Phys. Rev.\/} {\bf D 9} (1974) 608.
%
    \bibitem{Heller3} {\it}
K. Heller, In Proceeding of the 12th International Symposium on High-Energy
  Spin Physics, September 10 - 14, 1996. Amsterdam, The Netherlands.
Ed. by C.W. de Jager et al. World Sci., Singapore. (p. 23)  1996.
    \bibitem{Luk} {\it}
 K.B. Luk et al., {\em Phys. Rev. Lett.\/} {\bf 70} (1993) 900.
    \bibitem{Adamovich} {\it}
M.I. Adamovich et al., {\em Z. Phys.\/} {\bf A 350} (1995) 379.
%
    \bibitem{Polvado} {\it}
R.O. Polvado et al., {\em Phys. Rev. Lett.\/} {\bf 42} (1979) 1325.
%
    \bibitem{Kane} {\it}
 G.L. Kane, J. Pumplin and W. Repko,
 {\em Phys. Rev. Lett.\/} {\bf 41} (1978) 1689.

    \bibitem{Pondrom} {\it}
 L.G. Pondrom,  {\em Phys. Rep.\/} {\bf 122} (1985) 57. 

    \bibitem{Lach} {\it}
J.Lach,  Hyperon Polarization and Magnetic
   Moments. Preprint FERMILAB-Conf-93/381. 1993.
%
\bibitem{Soffer}
 J.Soffer, Is the riddle of the hyperon polarization solved?,
(Marseille, CPT). Preprint CPT-99-P-3898, Sep. 1999. 
Invited talk at Hyperon 99: Hyperon Physics Symposium, Batavia, Illinois, 27-29
Sep. 1999. p. 121;  hep-ph/9911373 (1999).
%
    \bibitem{Prep98} {\it}
  V.V. Abramov, A New Scaling for Single-Spin Asymmetry
         in Meson and Baryon  Hadroproduction. IHEP Preprint 98-84, Protvino,
         1998; hep-ph/0110152.
    \bibitem{EPJC} {\it}
  V.V. Abramov, {\em Eur. Phys. J.\/} {\bf C 14} (2000) 427;
 DOI 10.1007/s100529900355.
%
\bibitem{Liang1}
 Liang Zuo-tang and C.Boros, {\em Phys. Rev. Lett.\/} {\bf 79}  (1997) 3608.
%
%
    \bibitem{Bushnin} {\it}
   Yu. Bushnin  et al., {\em Phys. Lett.\/} {\bf  29} (1969) 48.
%
    \bibitem{Felix} {\it}
  J. Felix et al., {\em Phys. Rev. Lett.\/}  {\bf 76} (1996) 22.
%
    \bibitem{CIM1} {\it}
 R. Blankenbeckler and S.J. Brodsky, {\em Phys. Rev.\/} {\bf D 10} (1974) 2973.
%
    \bibitem{Qui} {\it}
 J. Qui and G. Sterman,Preprint ITP-SB-98-28; {\em Phys. Rev.\/} 
{\bf D 59} (1999) 014004; 
hep-ph/9806356. Single Transverse Spin Asymmetries in Hadronic Pion Production.
    \bibitem{Franklin} {\it}
 J. Franklin, {\em Phys. Rev.\/} {\bf 172}  (1968) 1807.
%

%
\bibitem{Boros1}
 C. Boros, Liang Zuo-tang, {\em Phys. Rev.\/} {\bf D 57} (1998) 4491.

%
\bibitem{Ryskin1}
 M.G. Ryskin,  Polarization phenomena and confinement forces,
 In Proc. of the Int. Conf. on Quark Confinement and the Hadron Spectrum,
Como, Italy, 20-24 June 1994. Edited by N.Brambilla and G.M.Prosperi.
River Edge, N.J., World Scientific, 1995, p. 261.
%

    \bibitem{Tyurin1} {\it}
  S.M. Troshin and N.E. Tyurin, {\em Phys. Rev.\/} {\bf D 55} (1997) 1265.

    \bibitem{Tyurin2} {\it}
  S.M. Troshin and N.E. Tyurin, {\em Phys. Rev.\/} {\bf D 52} (1995) 3862.
%
\bibitem{DeGrand1}
 T.A. DeGrand, H. Miettinen, {\em Phys. Rev.\/} {\bf D 24} (1981) 2419.
%
\bibitem{DeGrand2}
 T.A. DeGrand et al., {\em Phys. Rev.\/} {\bf D 32} (1985) 2445.
%
\bibitem{Thomas}
 L.T. Thomas, {\em Philos. Mag.\/} {\bf 3} (1927) 1.
%
\bibitem{Logunov}
A.A. Logunov, {\em On Tomas Precession.\/} IHEP preprint 98-85, Protvino, 1998.
%
\bibitem{Hadron_t}
 V.V. Abramov, {\em Yad. Fiz.\/} {\bf 44} (1986) 1318
 [{\em Sov. J. Nucl. Phys. \/} {\bf 44} (1986) 856].
%
\bibitem{PPB}
D.E.Groom et al., {\em Eur. Phys. J. \/} {\bf C 15} (2000) 1.
%
\bibitem{Ander}
 B. Anderson, G. Gustafson and G. Ingelman, {\em Phys. Lett.\/}
{\bf B 85} (1979) 417; {\em Phys. Rep.\/} {\bf 97} (1983) 31.
%
    \bibitem{RYSK}
 M.G. Ryskin, {\em Yad. Fiz.\/} {\bf 48} (1988) 1114
   [{\em Sov. J. Nucl. Phys. \/} {\bf 48} (1988) 708].
%
    \bibitem{UFN}
 S.M. Troshin, N.E. Tyurin, {\em Uspechi Fiz. Nauk\/} {\bf 164} (1994) 1073.
%
\end{thebibliography}
\end{document}